 \documentclass[final,5p,times,twocolumn]{elsarticle}

\usepackage{amssymb}
\usepackage{amsmath}

\usepackage{amsmath,amssymb,amsfonts}
\usepackage{tablefootnote}
\usepackage[flushleft]{threeparttable}
\usepackage[inline]{enumitem}
\usepackage{graphicx}
\usepackage{textcomp}
\usepackage{booktabs}
\usepackage{multirow}
\usepackage{subcaption}
\usepackage{url}
\usepackage[inline]{enumitem}
\usepackage{adjustbox}
\usepackage{algorithm}
\usepackage{float}

\usepackage[table,xcdraw]{xcolor}
\newcommand{\tcr}[1]{\textcolor{red}{#1}}

\definecolor{lightblue}{RGB}{230, 240, 250}
\usepackage{algpseudocode}
\usepackage{comment}
\usepackage{tcolorbox}

\tcbuselibrary{listingsutf8}

\newcommand{\multiLFramework}{\textsc{Multi-LF}}
\newcommand{\ddosh}{\textsc{DDoShield-IoT}}
\newcommand{\pcap}{PCAP}
\journal{}

\begin{document}

\begin{frontmatter}



\title{\multiLFramework: A Continuous Learning Framework for Real-Time Malicious Traffic Detection in Multi-Environment Networks}

\author[label1]{Furqan Rustam}
\author[label2]{Islam Obaidat}
\author[label1]{Anca Delia Jurcut}
\affiliation[label1]{organization={School of Computer Science, University College Dublin},
            addressline={},
            city={Dublin},
            postcode={D06WK27},
            country={Ireland}}

\affiliation[label2]{organization={Department of Computer Systems Technology, North Carolina A\&T State University},
            city={North Carolina},
            country={USA}}




\begin{abstract}
\emph{Multi-environment} (M-En) networks integrate diverse traffic sources, including \emph{Internet of Things} (IoT) and traditional computing systems, creating complex and evolving conditions for malicious traffic detection. Existing \emph{machine learning} (ML)-based approaches, typically trained on static single-domain datasets, often fail to generalize across heterogeneous network environments. To address this gap, we develop a realistic \emph{Docker–NS3-based testbed} that emulates both IoT and traditional traffic conditions, enabling the generation and capture of live, labeled network flows. The resulting \emph{M-En Dataset} combines this traffic with curated public \pcap~traces to provide comprehensive coverage of benign and malicious behaviors. Building on this foundation, we propose \emph{\multiLFramework}, a real-time continuous learning framework that combines a lightweight model (M1) for rapid detection with a deeper model (M2) for high-confidence refinement and adaptation. A confidence-based coordination mechanism enhances efficiency without compromising accuracy, while weight interpolation mitigates catastrophic forgetting during continuous updates. Features extracted at 1-second intervals capture fine-grained temporal patterns, enabling early recognition of evolving attack behaviors. Implemented and evaluated within the Docker–NS3 testbed on live traffic, \emph{\multiLFramework} achieves an accuracy of 0.999 while requiring human intervention for only 0.0026\% of packets, demonstrating its effectiveness and practicality for real-time malicious traffic detection in heterogeneous network environments.
\end{abstract}








\end{frontmatter}



\section{Introduction}
Modern networks face a broad spectrum of malicious activities that now extend far beyond volumetric denial of service floods to include data exfiltration, protocol abuse, and other adaptive threats~\cite{10380450}. These attacks originate from an increasingly diverse set of hosts, including cloud servers, enterprise endpoints, and resource-constrained \emph{Internet of Things} (IoT) devices, all interconnected in complex \emph{multi-environment} (M-En) networks~\cite{NAZIR2023101820}. The resulting variety in traffic patterns, timing characteristics, and device capabilities undermines traditional malicious detection methods that depend on static signatures or models tuned to a single operating context. The challenge is compounded by (i) the heterogeneity between IoT and traditional traffic, (ii) the absence of a realistic M-En dataset, (iii) the class imbalance that causes models to overfit the environment that dominates the training data ~\cite{rustam2024ai}, and (iv) the need for continuous monitoring with human expertise during online learning \cite{rustam2025adaptive}. Effective defense, therefore, requires a framework that can ingest live traffic from multiple sources (e.g., traditional systems and IoT devices) and continuously learn to adapt in real time to unseen traffic, all while imposing minimal overheads~\cite{ismail2021review,8299485}.

Researchers have addressed this emerging threat landscape by developing numerous \emph{machine learning} (ML) models that target specific deployment contexts \cite{MAHBOUBI2024104004}. Dedicated models have been trained for IoT traffic~\cite{devivo2024ddoshield}, for flows within software-defined networking fabrics~\cite{9527066,BHAYO2023106432}, and for conventional enterprise and backbone traffic (cf.,~\cite{najafimehr2023ddos}). These models achieve high accuracy within the intended domain; however, their performance degrades when encountering traffic that mixes protocols, spans diverse device classes, or carries attack patterns it has never seen during training. 
Similarly, several studies have explored M-En network security and addressed these challenges through various approaches. For instance, Rustam et al.~\cite{rustam2024famtds} propose a fully automated malicious traffic detection system that addresses the lack of M-En datasets by generating M-En traffic through a combined IoT and traditional network dataset (IoTID-20~\cite{IoTID20} and UNWNB-15~\cite{7348942}). They tackle traffic diversity by deploying a \textit{moth flame optimizer} (MFO) to find optimal weights of an ML algorithm for different types of network traffic. Similarly, another study by Rustam et al.~\cite{rustam2024malicious} employs an optimization approach to generate M-En traffic, which includes both IoT and traditional network traffic. They propose a self-diverse ensemble model by combining three variations of random forests optimized with a \textit{particle swarm optimizer} (PSO). Nonetheless, most current studies create M-En traffic by merging existing datasets and then training and testing their models on the resulting dataset, which may not capture real-time M-En network patterns. In addition, combining the feature sets of different networks (since each dataset provides distinct features) can lead to the loss of meaningful traffic patterns. Lastly, none of these studies employs continuous learning with proper human oversight, causing their models’ performance to degrade over time.

This paper addresses the challenges in securing M-En networks against malicious attacks in two main contributions. In the first contribution, we introduce a comprehensive approach to M-En dataset collection that utilizes the \ddosh~testbed~\cite{devivo2024ddoshield}, which leverages the integration of NS-3~\cite{riley2010ns} (a network simulator) with Docker~\cite{merkel2014docker} (a container-based virtualization technology) to host real-world binaries that communicate over a simulated network. This testbed enables us to generate realistic M-En traffic by running actual binaries in a simulated network, producing benign and malicious traffic from the traditional and IoT networks, respectively. To achieve a complete range of traffic types (i.e., to incorporate the missing benign IoT and malicious traditional traffic), we utilize external \pcap~datasets from the literature, ensuring our final dataset accurately reflects diverse and realistic M-En network conditions. Following our M-En dataset construction, we uniformly process all dataset packets through a consistent feature extraction pipeline, capturing both packet-level and time-based statistical features. This uniformity ensures that traffic from all sources can be directly compared and integrated. As a result, our M-En dataset provides a robust foundation for training ML models capable of detecting malicious attacks in M-En networks, offering improved reliability and generalizability compared to prior approaches that relied solely on merged feature sets without preserving raw packet-level information.

In the second contribution, we design and implement a continuous learning, real-time \emph{Multi-Level Framework} (\multiLFramework). \multiLFramework~combines two ML models, M1 and M2, and human (expert) intervention. Specifically, M1 is a lightweight model that is initially trained on a smaller, selected sample dataset. Furthermore, M1 is optimized for continuous learning with low computational costs, enabling adaptability to new traffic patterns and a faster detection rate with minimal resource requirements. M2, in contrast, is a more robust and complex model with higher accuracy, trained on a larger, more comprehensive dataset. In essence, \multiLFramework~operates as follows: when a new packet arrives, M1 attempts the first prediction; if its prediction confidence falls below a threshold, \multiLFramework~forwards the information about this packet to M2 for a second prediction attempt. Similarly, if M2 fails to reach a definitive prediction, \multiLFramework~escalate it for manual inspection by human experts. Then, \multiLFramework~updates M1 with the expert's label using a weight interpolation method where 75\% of M1's existing weights remain intact and 25\% come from a quick gradient update on the newly labeled sample. The updates to M1 introduce new knowledge while mitigating overfitting and catastrophic forgetting, which keeps M1 fast, adaptive, and accurate as traffic evolves.

We implement and deploy \multiLFramework~in \ddosh~to assess its performance against live traffic. Using this deployment, we achieve an accuracy of 0.999. Across fifteen streaming batches covering more than 115,795 packets, only three instances required human intervention, resulting in just 0.0026\% human involvement. Even with a high system depth, the latency rate remains unaffected. In addition, we measure resource usage and find that the lightweight M1 model maintains a throughput of 66,130 packets per second. 
In summary, the key contributions of this work are as follows.

\begin{itemize}
\item We propose \multiLFramework, a continuous learning, real-time framework that combines ML techniques with expert feedback for malicious attack detection in M-En networks. Code and dataset for reproducibility are available on GitHub\footnote{\url{https://github.com/furqanrustam/MultiLF}}.

\item  We construct an M-En traffic dataset by leveraging \ddosh~and integrating external \pcap~datasets. This dataset encompasses both IoT and traditional (non-IoT) traffic, including benign and malicious categories.

\item We propose transformer-based lightweight and complex model architectures, trained on a selective subset and a large corpus, respectively. Under \multiLFramework, these models achieve significant performance with high throughput while imposing minimal burden on human workload.

\item We test and validate \multiLFramework~in real-time scenarios by hosting its implementation in \ddosh. This process demonstrates that \multiLFramework~results in higher accuracy compared to recent approaches.

\item We collect a wide range of performance metrics (e.g., accuracy, precision, throughput, CO$_2$ emissions) and resource measurements (e.g., memory usage and CPU utilization) under real operating conditions.

\end{itemize}

The rest of this paper is organized as follows. Section~\ref{sec:relared_work} presents the related work in the problem domain. Section~\ref{sec:proposed_metho} describes the proposed methodology. Section~\ref{sec:eval_res_diss} discusses the evaluation and results, and Section~\ref{sec:coclu_future_work} provides the conclusion and future work.

\section{Related Work}
\label{sec:relared_work}
In this section, we discuss recent literature on malicious traffic detection for traditional, IoT, and M-En networks. Furthermore, we review the literature on continuous learning and identify gaps that our study addresses.

\subsection{Malicious Traffic Detection using Machine Learning}

Malicious traffic, e.g., \emph{Distributed Denial of Service} (DDoS) attacks, poses significant threats to network security. Many researchers have worked on efficiently mitigating these attacks. For example, Anley et al.~\cite{anley2024robust} propose an innovative approach that utilizes \emph{Convolutional Neural Networks} (CNNs), adaptive architectures, and transfer learning techniques to detect malicious traffic. Their method demonstrates robust detection capabilities across various attack categories and is validated on publicly available datasets such as CIC-DDoS2019~\cite{8888419}, CSE-CIC-IDS2018~\cite{Sharafaldin2018TowardGA}, UNSW-NB15~\cite{7348942}, and KDDCup'99~\cite{tavallaee2009detailed}. Their approach achieved accuracies of 93.62\%, 99.92\%, 99.84\%, and 98.99\% on each dataset, respectively. In a comparative study, Al-Eryani et al.~\cite{10485168} evaluate \emph{machine learning} (ML) algorithms using the CICDoS2019 dataset~\cite{8888419}, finding that \emph{Gradient Boosting} (GB) and XGBoost achieve high accuracy (GB: 99.99\%, XGBoost: 99.98\%) with minimal false alarms. 

Similarly, the study \cite{10948513} proposes a \textit{graph transformer-based autoencoder} (GTAE-IDS), an unsupervised packet-based graph intrusion detection framework designed for early and accurate anomaly detection. The system captures traffic and constructs sequential packet graphs, learning benign behavior using a graph autoencoder with transformer-based encoding to model contextual packet relationships. Unlike prior flow-based or supervised approaches, GTAE-IDS operates in near real time, requires no labeled data, and is resilient to unseen attacks. Experiments on CICIDS2017, CICIDS2018, and ACIIoT2023 show $>$98\% accuracy, even when observing only the first few packets of each communication. Another study \cite{11199889} introduces a scalable SDN-CFN network model integrated with a GNN–based anomaly detection framework (GNN-NAD). The proposed system combines \textit{software-defined networking} (SDN) and \textit{compute-first networking} (CFN) to enable intelligent, real-time traffic analysis. GNN-NAD fuses static, vulnerability-aware attack graphs with dynamic traffic features, providing a unified and context-rich view of network security. At its core, a GNN model (GSAGE) performs graph representation learning, and an RF classifier conducts final detection. The hybrid design (GSAGE + RF) achieves a significant accuracy of 99.69, even with limited samples from the CICIDS2017 dataset.

Persistent challenges in network defense stem from evolving attack vectors and the increasing complexity of network environments. In response, Zhao et al.~\cite{10380450} have developed DFNet, an approach that integrates advanced ML models with packet scheduling algorithms in the network data plane. This approach effectively forwards 99.93\% of victim-desired traffic during new attacks while incurring minimal overhead. Similarly, Singh et al.~\cite{singh2024novel} contribute to the detection and mitigation of different attacks in \emph{software-defined networking} (SDN) environments. Their work also presents a new dataset with over 1.7 million entries and employs two detection methods: Snort~\cite{waleed2022open} (an Intrusion Detection System) and eight different ML algorithms, including Ensemble Classifiers and a Hybrid Support Vector Machine-Random Forest (SVM-RF) classifier. The detection methods achieved 99.1\% accuracy. In addition, the authors suggested two strategies to mitigate malicious traffic: dropping illegitimate traffic and redirecting it.

\subsection{Malicious Traffic Detection in Traditional IP-based Network}
Traditional IP-based networks (networks that primarily handle non-IoT traffic using the Internet Protocol, distinguishing them from networks that may employ protocols like MQTT for IoT devices) have been extensively explored, resulting in numerous benchmark datasets and efficient security approaches.  In recent literature, several researchers have proposed methods to protect these networks from malicious actors. Talukder et al. \cite{talukder2024machine} present an ML-based network intrusion detection model that integrates \emph{Random Oversampling} (RO) to counter data imbalance, Stacking Feature Embedding derived from clustering results, and \emph{Principal Component Analysis} (PCA) for dimension reduction. Their model achieves exceptional accuracy rates: 99.59\% and 99.95\% with RF and \emph{Extra Trees} (ET) models on the UNSW-NB15 dataset~\cite{7348942}, 99.99\% on the CIC-IDS-2017 dataset~\cite{sharafaldin2019detailed}, and 99.94\% on the CIC-IDS-2018 dataset~\cite{Sharafaldin2018TowardGA} with DT and RF models, respectively.

Expanding the focus on network security, Casanova et al. [24] concentrate on transforming the CIRACIC-DoHBrw-2020 time-series dataset~\cite{9878102} for training deep learning models in network intrusion detection. Their approach includes a two-layer network classification strategy, distinguishing \emph{DNS over HTTPS} (DoH) from non-DoH traffic and further classifying DoH traffic into benign and malicious categories using a subset of 26 features and various types of \emph{Recurrent Neural Networks} (RNNs), such as \emph{Long Short-Term Memory} (LSTM), Bidirectional LSTM, \emph{Gated Recurrent Unit} (GRU), and Deep RNN. Bi-LSTM outperforms all other methods with 99\% accuracy, while GRU is the second-best performer. Hema et al. \cite{hema2023malicious} propose a novel feature selection metric, CorrAUC, and develop a new feature selection algorithm using a wrapper technique. Their approach enhances traffic flow classification accuracy, evaluated using the NSL-KDD dataset~\cite{tavallaee2009detailed} with three different ML algorithms, such as RF, LR, and KNN, achieving 99\%, 82\%, and 98\%, respectively. Babayigit et al.~\cite{10543240} propose the \emph{Queried Adaptive Random Forests} (QARF) method, an online active learning-based approach that combines adaptive RF with an adaptive margin sampling strategy. This method queries a small number of instances from unlabeled traffic streams to obtain training data. Experimental evaluations using the NSL-KDD dataset~\cite{tavallaee2009detailed} demonstrate that QARF achieves 98.20\% accuracy.

\subsection{Malicious Traffic Detection in IoT Networks}

IoT networks have attracted significant research interest due to their lower security mechanisms compared to traditional networks~\cite{omolara2022internet,LEFOANE2025104110}. For instance, Babayigit et al.~\cite{babayigit2024towards} propose a novel approach to IoT malicious traffic detection using multiple-domain learning. They used Edge-IIoTSet~\cite{9751703}, WUSTL-IIoT-2021~\cite{zolanvari2021wustl}, and X-IIoTID~\cite{9504604} datasets in the proposed approach. Furthermore, they employ an autoencoder for feature-space fusion to convert all datasets into a common feature space. Their hybrid deep learning model, combining CNN and GRU, achieves up to 97.68\% accuracy and improvements in transfer learning scenarios. Zhu et al.~\cite{10236538} present the \emph{Lightweight Knowledge Distillation Space-Time Neural Network} (LKD-STNN) model to address IoT security constraints by creating a compact model using knowledge distillation and adaptive temperature function dynamics. Their model achieves over 98\% accuracy on ToN-IoT~\cite{MOUSTAFA2021102994} and IoT-23~\cite{garcia2020iot23} datasets. 


Huo et al.~\cite{10536057} introduce LightGuard, a lightweight malicious traffic detection model for IoT. LightGuard utilizes \emph{lightweight residual block} (LRB) modules (inspired by ShuffleNetV2~\cite{vnwt}) and a novel ghost module for efficient feature map generation, achieving over 99.6\% accuracy across diverse datasets (Edge-IIoTset~\cite{9751703}, USTCTFC2016~\cite{lu2016ustc}, ToN-IoT~\cite{MOUSTAFA2021102994} and CIC-IoT~\cite{unb_iot_dataset} datasets) while maintaining low computational complexity. Babayigit et al.~\cite{babayigit2024towards} propose a multiple-domain learning framework to improve the reliability and generalization of DL models for Industrial IoT (IIoT) traffic classification. Their work integrates Edge-IIoTSet~\cite{9751703}, WUSTL-IIoT-2021~\cite{zolanvari2021wustl}, and X-IIoTID~\cite{9504604} datasets using an autoencoder for dimensionality harmonization and a modified locally linear embedding for statistical alignment. They use a hybrid DL model that combines CNNs and GRUs along with Bayesian optimization for hyperparameter tuning. Their work achieves 97.68\% accuracy, 97.70\% recall, 97.67\% precision, and 97.68\% F1-score for binary classification and improves to 97.80\% accuracy and 97.79\% F1-score with transfer learning.

\subsection{Continuous Learning for Malicious Traffic Detection}

Continuous learning is an approach in which a model learns continuously by incorporating new data without retraining from scratch~\cite{fan2019real}. This approach is especially useful for adapting to evolving data in dynamic environments. Continuous learning is also important in cybersecurity due to the evolving nature of attacks. Xu et al.~\cite{10472312} propose an approach that uses \emph{self-paced class incremental learning} (SPCIL). SPCIL leverages network traffic data to improve \emph{class incremental learning} (CIL), a deep learning technique that integrates new malware classes while preserving recognition of prior categories. SPCIL uses a loss function that combines sparse pairwise loss with sparse loss. Their experimental results demonstrate that SPCIL effectively identifies both existing and new malware classes. Compared to other incremental learning methods, SPCIL excels in performance and efficiency, with a minimal parameter count of 8.35 million, achieving accuracy rates of 89.61\%, 94.74\%, and 97.21\% in different test scenarios. Similarly, Ajjaj et al.~\cite{ajjaj2023incremental} present an approach to detect black hole attacks, an attack on the \emph{Ad hoc On-Demand Distance Vector} (AODV) routing protocol. They simulate realistic VANET scenarios using the \emph{Simulation of Urban Mobility} (SUMO)~\cite{krajzewicz2010traffic} and the Network Simulator (NS-3)~\cite{riley2010ns}. They evaluate the performance of two online incremental classifiers, \emph{Adaptive Random Forest} (ARF) and KNN, using metrics such as accuracy, recall, precision, and F1-score, as well as training and testing time. Results demonstrate that ARF successfully classifies and detects black hole nodes in VANETs, outperforming KNN in all performance measures.

Wang et al.~\cite{wang2023drnet} propose an incremental learning method for small-sample data to detect malicious traffic. Their method employs a pruning strategy to identify and remove redundant network structures, dynamically reallocating these resources based on the proposed measurement method according to the difficulty of the new class. Their approach ensures that the network can learn incrementally without overconsuming storage and computing resources. Additionally, the proposed method utilizes knowledge transfer to mitigate forgetting old classes, alleviating the burden of training large parameters with limited data. Their experimental results on multiple datasets outperform established baselines in classification accuracy while using 50\% less memory. In addition, Zhao et al.~\cite {zhao2024trident} propose Trident, a framework for detecting fine-grained unknown encrypted traffic. Trident transforms the identification of known and new classes into multiple independent one-class learning tasks. It comprises three modules: tSieve for traffic profiling, tScissors for outlier threshold determination, and tMagnifier for clustering. Their evaluations on four popular datasets show that Trident outperforms 16 other related works.

\subsection{Malicious Traffic Detection in Multi-Environment Networks}

Malicious traffic detection in M-En networks remains challenging due to the heterogeneous and complex nature of traffic patterns. M-En networks, where IoT and traditional IP-based traffic coexist, create challenges for the security system. Especially, different network types exhibit distinct traffic patterns, making it challenging to develop models that can effectively capture this diversity. To address this security concern, several studies propose different approaches to securing M-En. For example, Rustam et al.~\cite{rustam2024malicious} introduce a system that leverages the \emph{Synthetic Data Augmentation Technique} (S-DATE) and a PSO-based \emph{Diverse-Self Ensemble Model} (D-SEM) to enhance the detection of malicious activities across diverse network environments. Their approach is validated on a composite dataset integrating InSDN~\cite{9187858}, UNSW-NB15~\cite{7348942}, and IoTID-20~\cite{IoTID20}, achieving an accuracy of 98.9\%.

Building on this foundation, Rustam et al.~\cite{rustam2023securing} present a detection approach that combines a newly developed M-En traffic dataset with S-DATE to mitigate data imbalance and improve model training efficiency. This approach enhances the detection of malicious traffic, achieving a detection rate of 99.1\%. In advancing cybersecurity in M-En environments, Rustam et al.~\cite{rustam2024ai} develop ML models trained on AI-based traffic for the M-En dataset derived from benchmark datasets UNSW-NB15~\cite{7348942} and IoTID-20~\cite{IoTID20}. Their models counter both traditional and AI-based threats, achieving accuracy rates of 98.3\% for binary classification and 96.8\% for multi-class problems. Similarly, Indrasiri et al.~\cite{indrasiri2022malicious} propose an approach to detect malicious traffic in M-En networks through ensemble learning. By merging UNSW-NB15 and IoTID-20 datasets and reducing the feature set using \emph{Principal Component Analysis} (PCA), they develop a stacked ensemble model termed \emph{Extra Boosting Forest} (EBF). EBF enhances detection performance, achieving accuracy scores of 98.5\% and 98.4\% for binary and multi-class classifications, respectively. In addition, Zukaib et al.~\cite{zukaib2024meta} validate a framework designed for detecting cyberattacks in dynamic \emph{Internet of Medical Things} (IoMT) networks within M-En environments. Their approach integrates Federated Learning and Meta-learning within a multi-phase architecture, achieving an accuracy of 99.82\%.

\begin{table*}[h!]
\centering
\caption{Summary of Related Work}
\resizebox{0.9\textwidth}{!}{
\begin{threeparttable}
\begin{tabular}{lllp{50pt}p{50pt}p{60pt}p{50pt}p{50pt}lllll}
\hline
\textbf{Ref} & \textbf{Year} & \textbf{M-En} & \textbf{ML} & \textbf{DL} & \textbf{Dataset} & \textbf{Method} & \textbf{Results} & \textbf{RT} & \textbf{RE} & \textbf{CL} & \textbf{Early} & \textbf{Raw} \\ \hline

\rowcolor{lightblue}
\cite{indrasiri2022malicious} & 2022 & $\checkmark$ & ETC, GBM, RF & - & UNSWNB15, IoTID20 & EL, PCA & 98.50, 98.40 & $\otimes$ & $\checkmark$ & $\otimes$ & $\otimes$ & $\otimes$ \\

\cite{10485168} & 2023 & $\otimes$ & GBM, XGB & - & CICDoS2019 & - & 99.99, 99.98 & $\otimes$ & $\checkmark$ & $\otimes$ & $\otimes$ & $\otimes$ \\

\rowcolor{lightblue}
\cite{10380450} & 2023 & $\otimes$ & - & DFNet & SC & - & 99.93 & $\otimes$ & $\otimes$ & $\otimes$ & $\Box$ & $\otimes$ \\

\cite{casanova2023malicious} & 2023 & $\otimes$ & - & LSTM, BiLSTM, GRU, RNN & CIRACIC-DoHBrw2020 & TNC & 99.00 & $\otimes$ & $\otimes$ & $\otimes$ & $\otimes$ & $\otimes$ \\

\rowcolor{lightblue}
\cite{hema2023malicious} & 2023 & $\otimes$ & RF, LR, KNN & - & NSLKDD & CorrAUC, FS & RF: 99.00 & $\otimes$ & $\checkmark$ & $\otimes$ & $\Box$ & $\otimes$ \\

\cite{10543240} & 2023 & $\otimes$ & ARF & - & NSLKDD & OAL, AMS & 98.20 & $\otimes$ & $\otimes$ & $\checkmark$ & $\otimes$ & $\otimes$ \\

\rowcolor{lightblue}
\cite{10236538} & 2023 & $\otimes$ & - & LKD-STNN & ToNIoT, IoT23 & KD, ATFD & 98. & $\otimes$ & $\otimes$ & $\otimes$ & $\otimes$ & $\checkmark$ \\

\cite{10472312} & 2023 & $\otimes$ & - & SPCIL & USTCTFC2016 & AA & 89.61, 94.74, 97.21 & $\otimes$ & $\otimes$ & $\checkmark$ & $\otimes$ & $\otimes$ \\

\rowcolor{lightblue}
\cite{ajjaj2023incremental} & 2023 & $\otimes$ & ARF, KNN & - & VANET & AA & - & $\otimes$ & - & $\checkmark$ & $\otimes$ & $\otimes$ \\

\cite{wang2023drnet} & 2023 & $\otimes$ & - & 1D-CNN & - & PS, KT & 91.15 & - & - & $\checkmark$ & $\otimes$ & $\checkmark$ \\

\rowcolor{lightblue}
\cite{rustam2023securing} & 2023 & $\checkmark$ & ETC & - & IoTID20, UNSWNB15 & S-DATE & 99.10 & $\otimes$ & $\checkmark$ & $\otimes$ & $\otimes$ & $\otimes$ \\

\cite{anley2024robust} & 2024 & $\otimes$ & - & CNN, TL & CICDDoS2019, CICIDS2018, UNSWNB15, KDDCup99 & AA, CNN & 93.62, 99.92, 99.84, 98.99 & $\otimes$ & $\otimes$ & $\otimes$ & $\otimes$ & $\otimes$ \\

\rowcolor{lightblue}
\cite{singh2024novel} & 2024 & $\otimes$ & LR, SVM, NB, KNN, RF, EC, SVM-RF & ANN & SC & SVM-RF & 99.1 & $\otimes$ & $\checkmark$ & $\otimes$ & $\otimes$ & $\otimes$ \\

\cite{talukder2024machine} & 2024 & $\otimes$ & RF, ET, DT, XGB & - & UNSWNB15, CICIDS2017, CICIDS2018 & RO, SFE, PCA & 99.59, 99.99 & $\otimes$ & $\checkmark$ & $\otimes$ & $\otimes$ & $\otimes$ \\

\rowcolor{lightblue}
\cite{babayigit2024towards} & 2024 & $\otimes$ & - & CNN, GRU & EdgeIIoTSet, WUSTLIIoT2021, XIIoTID & MDL, AE, MLLE & 97.68 & $\otimes$ & $\otimes$ & $\otimes$ & $\otimes$ & $\otimes$ \\

\cite{zhao2024trident} & 2024 & $\otimes$ & - & AE, RNN, GNN & - & tSieve, tScissors& 96.71 & $\otimes$ & $\otimes$ & $\checkmark$ & $\otimes$ & $\Box$ \\

\rowcolor{lightblue}
\cite{rustam2024malicious} & 2024 & $\checkmark$ & RF & - & InSDN, UNSWNB15, IoTID20 & S-DATE, PSO & 98.90 & $\otimes$ & $\checkmark$ & $\otimes$ & $\otimes$ & $\otimes$ \\

\cite{rustam2024ai} & 2024 & $\checkmark$ & ETC & - & UNSWNB15, IoTID20 & AI-based MTC & 98.30 & $\otimes$ & $\checkmark$ & $\otimes$ & $\otimes$ & $\otimes$ \\

\rowcolor{lightblue}
\cite{zukaib2024meta} & 2024 & $\otimes$ & - & - & IoMT & FL, MeL & 99.82 & $\otimes$ & $\otimes$ & $\otimes$ & $\otimes$ & $\otimes$ \\

\cite{10948513} & 2025 & $\otimes$ & OCSVM, IF, HBOS, INNE & GAE, TE & CICIDS2017, CICIDS2018, ACIIoT2023 & GTAE-IDS & $>$98 & $\Box$ & $\otimes$ & $\otimes$ & $\checkmark$ & $\checkmark$ \\

\rowcolor{lightblue}
\cite{11199889} & 2025 & $\otimes$ & RF & GNN & CICIDS2017 & GSAGE+RF & 99.69  & $\otimes$ & $\otimes$ & $\otimes$ & $\otimes$ & $\otimes$ \\\hline\hline

Our & 2025 & $\checkmark$ & SGDC, PER, MNB, BNB & LSTM, LwT & SC M-En & AA, \multiLFramework & 99.99 & $\checkmark$ & $\checkmark$ & $\checkmark$ & $\checkmark$ & $\checkmark$ \\ \hline\hline
\end{tabular}

\begin{tablenotes}
\footnotesize

\item
\textbf{RTE}---(Real-Time Evaluation) whether the method evaluated in real-time;
\textbf{M-En}---(Multi-Environment) whether multiple network domains are used;
\textbf{RE}--- (Resources Efficiency);whether the method is resource efficient;
\textbf{CL}--- (Continuous Learning); whether the method get update over time with recent attacks;
\textbf{Raw}---Whether raw input data is used;
\textbf{Early}---Whether early traffic classification is used.
-----
\textbf{Acronyms}:
Transfer Learning (TL),  
Adaptive Architectures (AA),  
Self-Collected (SC), 
Stacking Feature Embedding (SFE),
Two-layer Network Classification  (TNC),  
Feature Selection (FS),  
Adaptive Random Forests (ARF),  
Online Active Learning  (OAL),  
Adaptive Margin Sampling (AMS),  
Multiple-Domain Learning (MDL), 
Autoencoder (AE), 
Modified Locally Linear Embedding (MLLE),
Knowledge Distillation (KD),
Adaptive Temperature Function Dynamics (ATFD),
Lightweight Residual Block (LRB),
Ghost Module (GM),
Self-Paced Class Incremental Learning (SPCIL),
Pruning Strategy (PS),
Knowledge Transfer (KT),
Ensemble Learning (EL),
Federated Learning (FL), 
Meta Learning (MeL),
Light-weighted Transformer (LwT), 
Graph-Autoencoder (G-AE),
Transformer-based Encoding (TE),
Isolation-based Anomaly Detection using Nearest Neighbor Ensembles (INNE), 
Histogram-based Outlier Score (HBOS),
Isolation forest (IF),
One-class Support Vector Machines (OCSVM),
-----
\textbf{Symbols}:
$\checkmark$~present, $\Box$~partial, $\otimes$~lacking;
``-'' indicates that the information is not provided in the paper.
\end{tablenotes}
\end{threeparttable}}
\end{table*}

\subsection{Gaps \& Limitations}

Despite significant advancements in malicious traffic detection using ML across various network environments, challenges remain. Most studies focus on specific network architectures, such as IoT \cite{10536057} or traditional networks \cite{talukder2024machine}. There is a lack of research in the M-En domain, with few studies \cite{rustam2023securing, zukaib2024meta} relying on synthetic combinations of existing benchmark datasets, as no realistic M-En datasets exist. This reliance on synthetic datasets may limit the generalizability of findings to real-world M-En networks. Most studies performed M-En experiments by merging datasets from different domains \cite{rustam2024famtds,rustam2024malicious1}, which can introduce unrealistic traffic patterns and lose critical semantics due to disrupted context and feature reduction \cite{el2009impact}. Moreover, adapting to continuously evolving threats, particularly advanced persistent threats utilizing sophisticated attack strategies, presents ongoing challenges. Some studies adopt continuous learning~\cite{wang2023drnet}, but they are domain-specific, focusing either on IoT or other networks without human oversights \cite{rustam2025adaptive}. Therefore, an approach is needed to handle M-En networks in real-time and update their knowledge with continuous learning. 

\section{Proposed Methodology}
\label{sec:proposed_metho}
\begin{figure}[h!]
\centering
\includegraphics[width=\columnwidth]{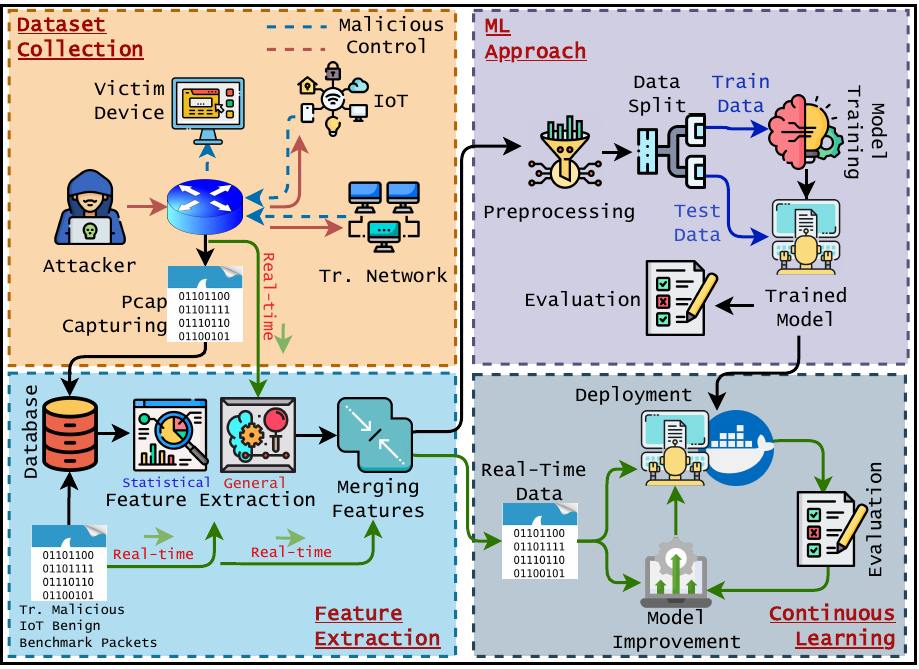}
\caption{Proposed Framework For M-En Malicious Traffic Detection}
\label{fig:overview}
\end{figure}

We introduce a real-time malicious traffic detection framework for M-En networks that combines machine learning with online, continuous learning, allowing the models to adapt to emerging traffic patterns as they emerge. Figure~\ref{fig:overview} outlines the workflow, which proceeds through four phases: (i) Dataset Collection, (ii) Feature Extraction, (iii) ML Approach, and (iv) Continuous Learning. We discuss the details of each phase in the following subsections.


\subsection{Phase 1: Data Collection}
\label{subsec:p1}
We construct a comprehensive dataset that encompasses both IoT and traditional (non-IoT) traffic, including both benign and malicious categories. To achieve this goal, we employ a two-pronged approach. First, we leverage \ddosh~\cite{devivo2024ddoshield}, a testbed that leverages the integration of Docker with NS-3 to host actual binaries that communicate over a simulated network, mimicking a realistic network environment. This setup enables the generation and capturing of real-world network traffic. We use \ddosh~to generate the two types of traffic it natively supports: benign traditional network traffic and malicious IoT network traffic. Second, to incorporate the other two missing traffic categories (benign IoT traffic and malicious traditional traffic), we integrate external \pcap~datasets (published in the literature) into \ddosh. Below, we discuss these two steps in detail, and then we discuss how we merge these different traffic sources in a unified manner.


\begin{enumerate}
    \item \textbf{Traffic Generation Using \ddosh:} \ddosh~supports the generation of certain traffic types. Specifically, it facilitates the creation of benign traditional traffic and malicious IoT traffic. For the benign traditional traffic, it generates different types of traffic (e.g., HTTP/HTTPS and video traffic) from legitimate services within Docker containers. For malicious IoT traffic, we utilize the Mirai malware~\cite{mohanty2018control} (a notorious IoT botnet) binaries hosted in a Docker container in \ddosh, from which we generate IoT-based attack traffic. In particular, we generate three Mirai-based attack classes: ACK flood, SYN flood, and UDP flood, which reflect key threats by IoT botnets in modern network environments~\cite{obaidat2023creating}.
    \item \textbf{Integrating External \pcap~Datasets:} While \ddosh~allows us to generate both benign traditional traffic and malicious IoT traffic, it does not natively support benign IoT traffic or malicious traditional traffic. To address these gaps, we incorporate external \pcap~datasets that have been proposed by other research works. This approach ensures our final dataset is both comprehensive and realistic. 
    
    For the benign IoT traffic, we use the MQTTIoTIDS2020 dataset\footnote{https://paperswithcode.com/dataset/mqtt-iot-ids2020}. The MQTT-IoT-IDS2020 dataset contains benign IoT network traces, focusing on IoT devices and services that utilize the MQTT protocol. It includes normal operations of IoT devices in a controlled environment, providing a clear baseline of legitimate IoT behavior. By integrating this \pcap~dataset, we incorporate authentic IoT benign traffic that closely represents everyday device activities.
    
    For the malicious traditional traffic, we use the CICDDoS2019 \pcap~dataset\footnote{https://www.unb.ca/cic/datasets/ddos-2019.html}. This dataset includes a variety of attacks captured in a realistic testbed environment. These attacks are carried out on traditional (non-IoT) services and include multiple attack vectors that target typical enterprise or data center hosts. Incorporating this dataset enhances the representativeness of our overall traffic corpus, ensuring that our traffic detection can identify malicious patterns in both IoT and traditional network segments.
\end{enumerate}

To combine these heterogeneous traffic sources, we deploy a Python sniffer running in a promiscuous Docker node inside \ddosh~so it sees every frame traversing the simulated network. At each injection interval, we select a \pcap~file from the external corpus and draw a random offset anywhere between its first and last packet. Rather than replaying traffic from that arbitrary byte position (which could split a session midstream), we step back to the first packet of the same flow (identified by the customary five-tuple of source IP, destination IP, source port, destination port, and transport-layer protocol~\cite{rfc3954}) and begin replaying from that clean boundary. The packets emitted from that point forward are merged with the live \ddosh~traffic and passed, without source labels, into the feature-extraction pipeline (discussed in~\S\ref{subsec:p2}). Because both the \pcap~file and the insertion point change every time, the resulting stream provides an unbiased, temporally coherent mix of IoT and traditional traffic segments.

All traffic is labeled according to established ground truths. Traffic generated within \ddosh~is inherently known to be malicious or benign based on controlled initiation of attacks or normal service activity. External \pcap~datasets come with predefined annotations, indicating which portions are malicious or benign. Thus, after merging, each packet (whether IoT or traditional) is labeled accordingly, ensuring clear distinctions among benign IoT, malicious IoT, benign traditional, and malicious traditional traffic.

\textbf{Attack Coverage: }Across the two malicious sources, our M‑En dataset spans fifteen distinct attack classes. \ddosh\ (Mirai) contributes three IoT‑centric floods (SYN, ACK, and UDP) that reflect the malware's most common attacks~\cite{obaidat2023creating}. The external CICDDoS2019 traces add twelve additional traditional attacks: NTP, PortScan, DNS, LDAP, MSSQL, NetBIOS, SNMP, SSDP, UDP, UDP‑Lag, WebDDoS, SYN, and TFTP floods. In addition, the MQTTIoTIDS2020 dataset provides two reconnaissance scenarios (Sparta SSH brute force and a UDP scan), which we include to capture low-volume probing behavior. This combination gives our corpus both high‑rate floods and short‑burst probing activity, providing broad coverage for M‑En detection research. Because \ddosh\ Docker and NS‑3 architecture lets us plug in new binaries or replay additional \pcap, future attack classes can be integrated with minimal effort, either by implementing the attack inside the testbed or by importing an annotated \pcap\ and following the same injection procedure described above.

\begin{tcolorbox}[colback=white, colframe=black!60, title={Attack Coverage},fonttitle=\bfseries,fontupper=\ttfamily\scriptsize]
SYN flood, ACK flood, and UDP flood, NTP, PortScan, DNS, LDAP, MSSQL, NetBIOS, SNMP, SSDP, UDP, UDP-Lag, WebDDoS, SYN, and TFTP flood, Sparta SSH brute-force, UDP scan
\end{tcolorbox}

\subsection{Phase 2: Feature Extraction}
\label{subsec:p2}



After capturing all M-En traffic, we apply a uniform feature extraction and aggregation methodology. This uniformity means that regardless of whether a packet is captured from the simulated environment or sourced from an external \pcap~dataset, the same set of features is extracted using identical definitions, thresholds, and time windows. By treating all traffic in this consistent manner, we eliminate biases that could arise from applying different techniques or settings to different portions of the data. As a result, this process presents a coherent feature space, making it possible to directly compare, combine, and evaluate M-En data from multiple origins using a single, unified analytic framework.

We categorize the extracted features into two types: general features (e.g., protocol types and payload sizes) and statistical features computed over fixed time intervals (e.g., the average packet size and the packet rate). General features are directly related to traffic characteristics, while statistical features are derived from analyzing general features over time to detect patterns indicative of malicious attacks.

The general features are essential for a comprehensive analysis and detection of malicious activities. The \textit{timestamp} records the precise time of packet capture, allowing for chronological analysis. \textit{Source and destination IP addresses (ip\_src, ip\_dst)} identify the communication endpoints, crucial for tracing attack origins and targets. The \textit{protocol} field specifies the transport protocol (e.g., TCP, UDP), aiding in protocol-specific analysis. \textit{Source and destination ports (src\_port, dst\_port)} indicate the application-level endpoints, useful for identifying targeted services. The presence of \textit{TCP and UDP flags (tcp\_flag, udp\_flag)} denotes the protocol used, while \emph{Time to Live} (TTL) reveals the packet's hop count, indicating network topology and potential anomalies. Flags such as \textit{ACK, SYN, FIN, PSH, URG, and RST} in TCP packets provide insight into connection states and potential malicious behaviors, including SYN flooding or connection resets. \textit{Sequence and acknowledgment numbers} are vital for reconstructing TCP sessions and detecting session hijacking or manipulation. Finally, the \textit{packet size} and \textit{payload size} metrics help identify unusual packet structures and potential payload-based attacks. Together, these features provide a detailed view of network traffic, enabling effective identification and mitigation of diverse cyber threats.

\begin{tcolorbox}[colback=white, colframe=black!60, title=General Feature, fonttitle=\bfseries,fontupper=\ttfamily\scriptsize]
\begin{verbatim}
'Timestamp', 'Source', 'Destination', 'Protocol', 'SrcPort', 
'DstPort', 'TCP', 'UDP', 'TTL', 'ACK', 'SYN', 'FIN', 
'PSH', 'URG', 'RST', 'SequenceNumber', 'PacketSize', 
'AcknowledgmentNumber', 'PayloadSize'
\end{verbatim}
\end{tcolorbox}

Despite the value of general features (per-packet descriptors), they alone cannot flag malicious activity reliably, because many network attacks reveal themselves through temporal patterns (sustained bursts of traffic, rapid-fire connection attempts, or repeated request sequences) that only emerge when packets are viewed in aggregate~\cite{fouladi2013frequency}. Therefore, we also extract statistical features over fixed windows to expose such patterns and anomalies. These 24 statistical features are crucial for identifying the diverse range of attacks. Our feature extraction pipeline is modular, allowing for the easy integration of additional features whenever new threats or protocols emerge. Below, we list the selected statistical features and explain the rationale behind the key thresholds; the full calculation details are provided in Appendix~\ref{app1}.

\begin{tcolorbox}[colback=white, colframe=black!60, title=Statistical Feature, fonttitle=\bfseries,fontupper=\ttfamily\scriptsize ]
\begin{verbatim}
'Packet Count', 'Destination Port Entropy', 'Most Frequent 
Source Port', 'Most Frequent Destination Port',
'Short-lived Connections', 'Repeated Connection Attempts',
'Network Scanning Activity', 'Flow Rate, Source Entropy', 
'Connection Errors~(RST)', 'Most Frequent Packet
Size Frequency', 'Abnormal Packet Size Frequency', 'Sequence
Number Variance', 'SYN Frequency', 'ACK Frequency',
'TCP Frequency', 'UDP Frequency', 'Packet Size Variability', 
'Average Packet Size', 'Average Payload Size'
\end{verbatim}
\end{tcolorbox}



\textbf{Thresholds \& Parameters For Simulation: }In our approach, we use specific thresholds and parameters to calculate these statistical features as shown in Table \ref{tab:simulation_parameters}. We set the \textit{processing interval} to 1\,s, a value that strikes a balance between latency and statistical stability; in a preliminary sweep of 100\, ms–2\,s windows, 1\,s produced the highest F\textsubscript{1} score while keeping detection delay sub-second. We adopt 1500\,B as the \textit{abnormal size threshold} because it aligns with the default IP \emph{maximum-transmission-unit} (MTU) on typical networks~\cite{rfc4821,rfc5948}; packets larger than this value ordinarily require fragmentation, so their presence is a useful indicator of atypical or crafted traffic, both of which warrant closer inspection. The \textit{port-frequency threshold} is set to five occurrences per window, an empirically derived cutoff that isolates the heavy hitters responsible for a large portion of connection attempts in our training set. Finally, we label any TCP flow containing fewer than five packets as a short‑lived connection. Previous work treats such ultra‑short sessions as abnormal traffic that ends before any meaningful application‑layer exchange can occur~\cite{kotenko2017parallel,jose2021towards}.

\begin{table}[h]
\centering
\caption{Thresholds and Parameters Used in Simulation}
\resizebox{\columnwidth}{!}{
\begin{threeparttable}
\begin{tabular}{lp{23mm}p{65mm}}
\hline
\textbf{Parameter} & \textbf{Value} & \textbf{Description} \\
\hline
     \rowcolor{lightblue}
PI & 1s & Balances latency and statistical stability; yielded the highest F1-score in a 100~ms–2~s sweep. \\
APST & 1500 B & Corresponds to the typical IP MTU; packets exceeding this value may indicate fragmentation or crafted traffic. \\
\rowcolor{lightblue}
PFT & 5 occurrences/ window & Empirical cutoff isolating frequently targeted ports responsible for most connection attempts. \\
SLCT & $<$ 5 packets/flow & Identifies ultra-short TCP sessions considered abnormal or incomplete exchanges. \\
\hline
\end{tabular}
\begin{tablenotes}
\scriptsize
\item
\textbf{PI}-- Processing Interval; 
\textbf{APST}-- Abnormal Packet Size Threshold;
\textbf{PFT}--- Port-Frequency Threshold;
\textbf{SLCT}--- Short-Lived Connection Threshold;
\end{tablenotes}
\end{threeparttable}
}
\label{tab:simulation_parameters}
\end{table}

These choices follow best practice in flow-analysis literature (e.g., NetFlow guidelines~\cite{rfc3954}) and are validated on a held-out subset of our dataset; nevertheless, all of these parameters remain tunable knobs in our implementation. Practitioners can re-estimate them with a grid search or Bayesian optimization to suit higher-speed links or encrypted-traffic mixes. By combining the general features with these carefully parameterized statistical features, we distinguish normal from malicious traffic more effectively. 

\subsection{Phase 3: ML Approach}
\label{subsec:p3}
\begin{figure}[h!]
    \centering
    \includegraphics[width=\columnwidth]{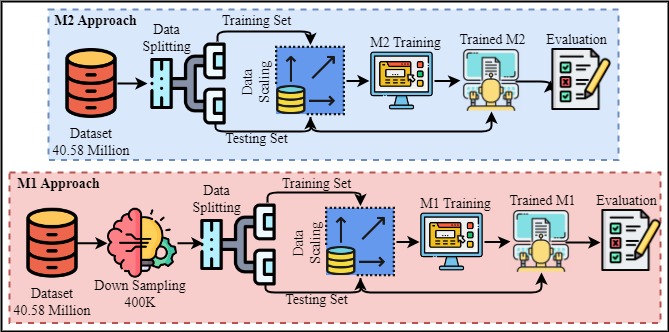}
    \caption{M1 \& M2 Models Training Approaches}
    \label{fig2}
\end{figure}

In the ML approach, we employ several state-of-the-art methods for detecting malicious traffic. We train two models, M1 and M2. M1 is a lightweight model trained on selected samples after preprocessing, while M2 is a complex model trained on a large dataset, as shown in Figure \ref{fig2}. M1 directly processes live traffic, and if the confidence value of the detection is below a threshold, the decision is passed to M2, which attempts to perform the prediction (and retrain M1 based on its prediction).

\subsubsection{\textbf{Baseline M1 \& M2:}} In the M1 training strategy, we begin by selecting a balanced subset of the data, 100,000 samples per class, totaling 400,000 samples, to create a lightweight yet representative dataset. Initially, we train models using randomly selected samples. To refine the dataset, we then apply K-Means clustering \cite{vzalik2008efficient} to the standardized features, selecting the top 100,000 samples per class that are closest to their respective cluster centers. This process ensures each class is represented by its most central and diverse instances. After selection, we scale the features using the MinMax method and split the dataset into 75\% training and 25\% testing. Then, we train four models: \emph{stochastic gradient descent classifier} (SGDC), PER, \emph{multinomial naive Bayes} (MNB), and \emph{Bernoulli naive Bayes} (BNB). We train all models using the partial\_fit function to allow continual learning. We evaluate the model's performance using accuracy, precision, recall, and F1 score, along with 10-fold cross-validation for robustness. We select the model hyperparameters based on literature~\cite{rustam2024ai,rustam2024malicious,gada2021automated} and tune them within specific ranges and values using a trial-and-error approach. We use BNB and MNB with their default settings, which are effective due to their simplicity and low sensitivity to hyperparameters~\cite{ismail2022evaluation}. The following sections provide further analysis of each model and its suitability for the M-En malicious traffic detection task.

\begin{table}[h!]
    \centering
        \caption{Hyperparameter Settings For Baseline M1 Models}
    \begin{tabular}{llll}
        \hline
        \textbf{Model} & \textbf{Hyperparameter} & \textbf{Value} & \textbf{Tuning Range}\\\hline
            \rowcolor{lightblue}
        SGDC & loss & log\_loss & - \\
            \rowcolor{lightblue}
                      & max\_iter & 1000 & [200 to 1500] \\
                          \rowcolor{lightblue}
                      & tol & 1e-3 & - \\
                          \rowcolor{lightblue}
                      & random\_state & 42 & [0, 42] \\

        PER & max\_iter & 1000 & [200 to 1500]\\
                   & tol & 1e-3 & - \\
                   & random\_state & 42 & [0, 42] \\

        \rowcolor{lightblue}
        MNB & Default & Default & - \\

        BNB & Default & Default & - \\
        \hline
    \end{tabular}
    \label{tab:hyperparameters1}
\end{table}

\textbf{\(\Rightarrow\)SGDC:}
It is a linear classifier that optimizes a linear model using stochastic gradient descent. We use it for malicious traffic detection in M-En because it is efficient for large-scale and sparse datasets. It also supports continuous learning, which makes it suitable for our study \cite{angelopoulos2021impact}. It supports various loss functions; however, we use log loss, which helps us work with prediction probabilities to measure model confidence. The SGDC models the decision function as a linear combination of the input features as:
\(
    z = \mathbf{w} \cdot \mathbf{x} + b = w_1 x_1 + w_2 x_2 + \ldots + w_n x_n + b
\)
where, \(\mathbf{w} = [w_1, w_2, \ldots, w_n]\) is the weight vector, \(b\) is the bias term, \(\mathbf{x}\) is the input vector. Depending on the chosen loss function (log in our case), the SDGC optimizes the model using the gradient of the loss function, which can be described as:

\begin{equation}
    L(\mathbf{w}, b) = -\log \left( \frac{1}{1 + \exp(-y (\mathbf{w} \cdot \mathbf{x} + b))} \right)
\end{equation}

Then SGDC updates weights iteratively, which can be described as:
\(
    \mathbf{w} \leftarrow \mathbf{w} - \eta \nabla L(\mathbf{w}, b, \mathbf{x}_i, y_i)
\);
\(
    b \leftarrow b - \eta \frac{\partial L(\mathbf{w}, b, \mathbf{x}_i, y_i)}{\partial b}
\)
where, \(\eta\) is the learning rate, \(\nabla L\) is the gradient of the loss function.

\textbf{\(\Rightarrow\)PER:}
It is a linear classifier that makes predictions based on a simple threshold function \cite{sudqi2019lightweight}. It is used for binary classification tasks and can be extended to multiclass classification using the \emph{one-vs-rest} (OvR) scheme, as in our M-En malicious traffic detection. It is also effective when data is linearly separable, which is very suitable for our case, as our dataset is highly linearly separable. PER in sci-kit-learn also provides a continuous learning function, so we use it as M1. The PER classifier aims to find a hyperplane that separates the classes. Given an input vector \(\mathbf{x} = [x_1, x_2, \ldots, x_n]\), the PER computes the output \(y\) using:
\(
    z = \mathbf{w} \cdot \mathbf{x} + b = w_1 x_1 + w_2 x_2 + \ldots + w_n x_n + b
\)
where, \(\mathbf{w} = [w_1, w_2, \ldots, w_n]\) is the weight vector, \(b\) is the bias term, \(\mathbf{x}\) is the input vector. The output \(y\) is then determined by applying the activation function:

\begin{equation}
    y = \begin{cases} 
1 & \text{if } z \geq 0 \\
-1 & \text{if } z < 0 
\end{cases}
\end{equation}

\textbf{\(\Rightarrow\)MNB:}
It is suitable for large datasets where fast computation is needed, as in our case \cite{panda2010discriminative}. It also provides the \textit{`partial\_fit`} function for continuous learning, which we use in our approach as the M1 model in comparison with others \cite{abdulboriy2024incremental}. It estimates the probability of a class given the feature values by combining the prior probability of the class and the likelihood of the observed features given the class using Bayes' theorem:

\begin{equation}
    P(C_k | \mathbf{x}) = \frac{P(C_k) \cdot P(\mathbf{x} | C_k)}{P(\mathbf{x})}
\end{equation}

where, \(P(C_k | \mathbf{x})\) is the posterior probability of class \(C_k\), \(P(C_k)\) is the prior probability of class \(C_k\), \(P(\mathbf{x} | C_k)\) is the likelihood of \(\mathbf{x}\) given \(C_k\), \(P(\mathbf{x})\) is the marginal likelihood of \(\mathbf{x}\). The likelihood is estimated based on the frequency of features, and parameters for MNB are estimated using Laplace smoothing.

\textbf{\(\Rightarrow\)GNB:}
We use it because network traffic features such as packet sizes, inter-arrival times, and other statistical metrics are continuous. GNB is well-suited for continuous data as it assumes that the features follow a Gaussian (normal) distribution \cite{meerja2021gaussian}. It also uses Bayes' theorem like MNB:

\begin{equation}
    P(C_k | \mathbf{x}) = \frac{P(C_k) \cdot P(\mathbf{x} | C_k)}{P(\mathbf{x})}
\end{equation}

where, \(P(C_k | \mathbf{x})\) is the posterior probability of class \(C_k\), \(P(C_k)\) is the prior probability of class \(C_k\), \(P(\mathbf{x} | C_k)\) is the likelihood of \(\mathbf{x}\) given \(C_k\), \(P(\mathbf{x})\) is the marginal likelihood of \(\mathbf{x}\). In the GNB model, the likelihood is assumed to be normally distributed:

\begin{equation}
    P(x_i | C_k) = \frac{1}{\sqrt{2\pi\sigma_{C_k, i}^2}} \exp\left(-\frac{(x_i - \mu_{C_k, i})^2}{2\sigma_{C_k, i}^2}\right)
\end{equation}

where, \(\mu_{C_k, i}\) is the mean of \(x_i\) in \(C_k\), \(\sigma_{C_k, i}^2\) is the variance of \(x_i\) in \(C_k\). Further, parameters are estimated using maximum likelihood estimation in GNB.





\textbf{\(\Rightarrow\)Baseline M2:} In the M2 training approach, models were trained on the full dataset of 40 million samples. After applying data scaling, the dataset was split into 75\% for training and 25\% for testing. We trained various models, including MLP~\cite{setitra2023optimized}, LR~\cite{chalichalamala2023logistic}, SGDC~\cite{azimjonov2024stochastic}, BNB~\cite{sharmila2019intrusion}, GNB~\cite{nugroho2021ensemble}, random forest (RF)~\cite{hasan2016feature}, AdaBoost (ADA)~\cite{hu2008adaboost}, support vector classifier (SVC)~\cite{khan2007new}, and k-nearest neighbors (KNN)~\cite{govindarajan2009intrusion}. Trained M1 and M2 models were saved using the \texttt{pickle} library with a \texttt{.pkl} extension for real-time testing. As with M1, M2 models were deployed with their best hyperparameter configurations, determined through literature review and iterative tuning~\cite{ali2023hyperparameter,rustam2024famtds,gada2021automated}. Table~\ref{tab:hyperparametersM2} lists the hyperparameter ranges and values used. Models such as BNB, GNB, and MNB were deployed using default settings due to their robustness and minimal sensitivity to hyperparameters.


\begin{table}[h!]
    \centering
        \caption{Hyperparameter Settings For Baseline M2 Models}
    \begin{tabular}{llll}
        \hline
        \textbf{Model} & \textbf{Hyperparameter} & \textbf{Value} & \textbf{Tuning Range}\\
        \hline
                \rowcolor{lightblue}
        LR & max\_iter & 1000 & [200 to 1500 ] \\
                \rowcolor{lightblue}
                           & random\_state & 42 & [0, 42] \\

        SGDC & loss & log & - \\
                      & max\_iter & 1000 & [200 to 1500 ] \\
                      & tol & 1e-3 & - \\
                      & random\_state & 42 & [0, 42] \\
        \rowcolor{lightblue}
        PER & max\_iter & 1000 & [200 to 1500 ] \\
                \rowcolor{lightblue}
                   & tol & 1e-3 & - \\
                           \rowcolor{lightblue}
                   & random\_state & 42 & [0, 42] \\

        RF & n\_estimators & 100 & [10 to 300] \\
                               & random\_state & 42 & [0, 42]\\
           &       max\_depth  & 80  & [2 to 100]    \\ 
                   \rowcolor{lightblue}
        ADA & n\_estimators & 100 & [10 to 300] \\
                \rowcolor{lightblue}
                           & random\_state & 42 & [0, 42] \\
                                   \rowcolor{lightblue}
            &       max\_depth  & 80  & [2 to 100]\\
            
        SVC & probability & True & -\\
            & random\_state & 42 & [0, 42] \\
        \rowcolor{lightblue}
        KNN & n\_neighbors & 5 & [3 to 30] \\

        MLP & hidden\_layer\_sizes & (100,) & [50 to 500] \\
                      & max\_iter & 1000 & [200 to 1500]  \\
                      & random\_state & 42 & [0, 42]\\
        \rowcolor{lightblue}
        MNB & Default & Default & - \\

        BNB & Default & Default & -  \\
        \rowcolor{lightblue}
        GNB & Default & Default & -  \\
        \hline
    \end{tabular}
    \label{tab:hyperparametersM2}
\end{table}

\subsubsection{\textbf{Proposed DL-Based M1 \& M2:}} 
In this study, we propose two DL-based architectures for M1 and M2: a lightweight Transformer (\textit{T1} $\rightarrow$ M1), a complex Transformer (\textit{T2} $\rightarrow$ M2), a lightweight LSTM (\textit{L1} $\rightarrow$ M1), and a complex LSTM (\textit{L2} $\rightarrow$ M2). We evaluate these models against several baselines. The key differences between the lightweight and complex architectures lie in the size of the training datasets and the number of layers in each architecture.

\begin{figure}[h!]
    \centering
    \includegraphics[width=1.05\columnwidth]{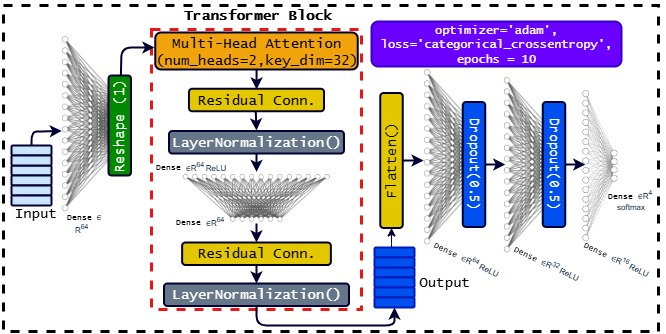}
    \caption{Proposed DL-based T2 Model Architecture }
    \label{figT2}
\end{figure}

\textbf{\(\Rightarrow\)Transformer-based Approach:} The \textit{T1} architecture consists of a single dense layer followed by a transformer block with multi-head self-attention and a simple feedforward network. After flattening the attention output, it uses one dense layer before the final softmax classification layer. This design ensures minimal computational overhead, making it suitable for resource-constrained or real-time environments. In contrast, the \textit{T2} architecture includes an additional dense layer at the input, a deeper classification head with three fully connected layers (64~$\rightarrow$~32~$\rightarrow$~16), and two dropout layers with a rate of 0.5 to mitigate overfitting. Both models share the same transformer block structure, but the complex version significantly increases representational capacity and depth, enabling improved learning from complex data patterns at the cost of increased training time and memory usage. Figure~\ref{figT2} illustrates the architecture configuration of \textit{T2}, the complex version of \textit{T1}.

\textbf{\(\Rightarrow\)LSTM-based Approach:}
Similarly, the \textit{L1} architecture consists of two LSTM layers: the first with 64 units and the second with 32 units. The final LSTM output is passed through a dense layer with 64 units and a ReLU activation, followed by a dropout layer with a rate of 0.3 before the final softmax classification layer. This configuration is lightweight and well-suited for real-time or resource-constrained environments. In contrast, the \textit{L2} architecture is deeper and more regularized. It begins with an LSTM layer with 64 units, followed by another LSTM layer with 32 units (both with \texttt{return\_sequences=True}), and a third LSTM layer with 16 units. Between the LSTM layers, dropout layers with a rate of 0.5 are applied to reduce overfitting. The final recurrent output is passed through two dense layers (64 and 32 units, respectively), with an additional dropout of 0.3 before the final softmax output. This deeper and more complex architecture is designed to better capture sequential dependencies in more challenging or high-volume traffic data scenarios.

All models in this study were compiled using the \texttt{Adam} optimizer and the \texttt{categorical\_crossentropy} loss function, which is suitable for multi-class classification tasks. The models were trained for 10 epochs with a batch size of 128. These settings were consistently applied across all Transformer- and LSTM-based architectures.
\color{black}

\subsection{Phase 4: \multiLFramework~Framework}
\label{subse:p4}



\begin{figure}[h!]
    \centering
    \includegraphics[width=1\columnwidth]{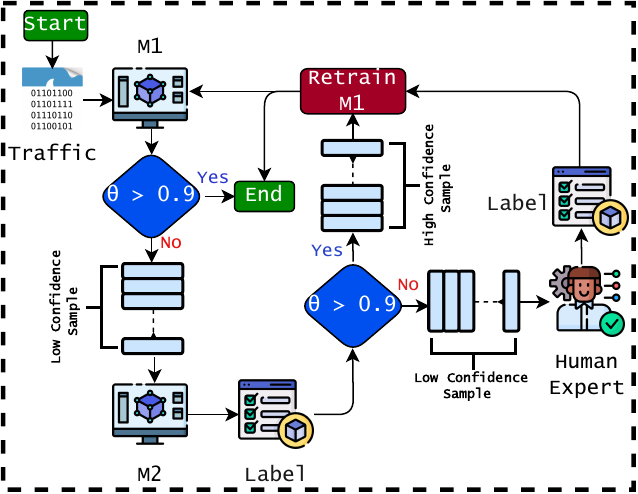}
    \caption{Overview of \multiLFramework~Flow Diagram}
    \label{fig1}
\end{figure}

In the \multiLFramework, we utilize two models: M1 (a lightweight model trained on a small subset of the dataset) and M2 (a more complex model trained on the full dataset). Initially, incoming data is passed to M1 for traffic label prediction. If M1's prediction confidence is 90\%, it proceeds to the next sample or terminates the process. If M1's confidence is below 90\%, the control shifts to M2, which then performs its own prediction. If M2's confidence exceeds 90\%, M1 is retrained using the predicted label from M2 through a continuous learning mechanism. To mitigate the catastrophic forgetting problem during this retraining, we employ a weight interpolation technique, where the updated weights \( \mathbf{w}_{\text{new}} \) are computed as:
\(
\mathbf{w}_{\text{new}} = \alpha \mathbf{w}_{\text{updated}} + (1 - \alpha) \mathbf{w}_{\text{old}},
\)with \( \alpha = 0.75 \), balancing between the new and old weights. 
The parameter $\alpha$ controls the trade-off between model adaptability and knowledge retention. In this work, $\alpha = 0.75$ is selected to preserve 75\% of prior information while allowing adaptation to new traffic patterns \cite{rustam2025adaptive}. This value was determined empirically through a series of trials with $\alpha \in {0.25, 0.50, 0.75, 1.0}$. Lower values (0.25 and 0.50) caused the model to overfit recent samples and forget previously learned patterns, whereas $\alpha = 1.0$ completely inhibited adaptation to new data. Since our continuous learning setup receives relatively small batches of incoming traffic, $\alpha = 0.75$ provided the most stable and consistent performance, and this configuration is reported in our results. 
Further, if M2's confidence is also below 90\%, control is transferred to a human expert, who labels the traffic sample. M1 is then retrained using the new data and the expert-provided label. This approach enables continuous learning, expert interaction, and ongoing performance enhancement of both models, as illustrated in Figure~\ref{fig1}.

Algorithm~\ref{algo1} describes the continuous learning process. 
Let \( \mathcal{T} \) denote the incoming traffic stream, and \( x \) represent a current traffic sample. The models used are \( M_1 \) (a lightweight model) and \( M_2 \) (a complex model), trained on datasets \( \mathcal{D}_s \) (a small subset) and \( \mathcal{D} \) (the full dataset), respectively. During prediction, \( M_1 \) and \( M_2 \) produce predicted labels \( \hat{y}_1 \) and \( \hat{y}_2 \), with associated confidence scores \( \gamma_1 \) and \( \gamma_2 \). If confidence is low, a human expert provides a label \( \hat{y}_h \). The final output label is denoted by \( \hat{y} \), and the confidence threshold for decision-making is set as \( \theta = 0.9 \).


    
        

\begin{algorithm}[h!]
\caption{Proposed \multiLFramework~Algorithm}
\label{algo1}
\begin{algorithmic}[1]
\State \textbf{Input:} Dataset \( \mathcal{D} \), Confidence Threshold \( \theta = 0.9 \)
\State \textbf{Output:} Predicted label \( \hat{y} \) for incoming traffic stream \( \mathcal{T} \)
\State \textbf{Initialize:} Train model \( M_1 \) on subset \( \mathcal{D}_s \) (400K samples using K-Means); train model \( M_2 \) on full dataset \( \mathcal{D} \) (40.58M samples)

\While{\( \mathcal{T} \) is not empty}
    \State \( x \gets \text{next\_sample}(\mathcal{T}) \)
    \State \( \hat{y}_1, \gamma_1 \gets M_1.\text{predict}(x) \)

    \If{\( \gamma_1 \gets \theta \)}
        \State \( \hat{y} \gets \hat{y}_1 \)
    \Else
        \State \( \hat{y}_2, \gamma_2 \gets M_2.\text{predict}(x) \)

        \If{\( \gamma_2 > \theta \)}
            \State \( \hat{y} \gets \hat{y}_2 \)
            \State \( M_1.\text{retrain}(x, \hat{y}_2) \)
        \Else
            \State \( \hat{y}_h \gets \text{get\_human\_label}(x) \)
            \State \( \hat{y} \gets \hat{y}_h \)
            \State \( M_1.\text{retrain}(x, \hat{y}_h) \) \Comment{Periodic human updates}
        \EndIf
    \EndIf
    \State Store or use \( \hat{y} \)
\EndWhile
\end{algorithmic}
\end{algorithm}

\newpage

\section{Evaluation and Results Discussion}
\label{sec:eval_res_diss}

We conduct all training and evaluation on an Intel Core i7-11800H host equipped with 64 GB RAM and a 1 TB SSD. The experiments run inside an Ubuntu 24.04 virtual machine (VMware Workstation) with Python 3.12 and scikit-learn 1.6. To emulate realistic conditions, we replay mixed IoT and traditional traffic through \ddosh~and classify this traffic on a promiscuous Docker node that captures the entire packet stream. We report accuracy, precision, recall, F1 score, runtime, memory footprint, and estimated CO$_2$ emissions (via CodeCarbon~\cite{courty2024codecarbon}) for experiments.



\subsection{Baseline Lightweight M1 Models Results}

Table~\ref{tab:classification_reportM1} shows results for the baseline M1 models, trained on a 400,000-instance sample using partial fit, with and without min-max scaling. Without scaling, the performance of the models is poor, except for BNB, which achieves an accuracy score of 1.000, and MNB, which reaches an accuracy score of 0.972. In contrast, both SGDC and PER have a very low accuracy scale. These results occur since SGDC and PER are linear models that are sensitive to the scale of features. When features have different scales, the gradient descent optimization process becomes inefficient, leading to poor convergence. The models struggle to learn effectively from unscaled data, resulting in poor performance across most classes. However, the performance of all models with scaling is excellent in all evaluation metrics, with all models achieving scores of 1.00. Scaling ensures that all features contribute proportionately to the model, enhancing the training efficiency and effectiveness of SGDC and PER classifiers. Additionally, scaling normalizes the feature space, leading to more balanced weight updates and improved model performance. For instance, if packet sizes range from 0 to 10,000 and another feature, such as traffic rate, ranges from 0 to 1, the larger feature disproportionately influences the model's parameters, leading to suboptimal performance. Here, scaling mitigates this problem by normalizing the features. Similarly, the MNB and BNB models assume feature independence and use probabilities based on the occurrence of features. Without scaling, the probability estimates can become skewed, affecting the model's ability to correctly classify instances. 

\begin{table}[h!]
\centering
\caption{Baseline M1 Models Results Using M-En Data}
\label{tab:classification_reportM1}
\begin{adjustbox}{width=\columnwidth}
\begin{tabular}{llllll}
\hline
\multicolumn{6}{c}{\textbf{With Scaling}}\\ \hline
\textbf{Model}   & \textbf{Acc}     & \textbf{Class} & \textbf{Pre} & \textbf{Rec} & \textbf{F1} \\ \hline
\rowcolor{lightblue}
\multirow{5}{*}{\cellcolor{lightblue}SGDC} &  \multirow{5}{*}{\cellcolor{lightblue}1.000}  &IoT.B & 1.00 & 1.00 & 1.00 \\ 
                            &  & IoT.M & 1.00 & 1.00 & 1.00 \\ 
                        
                            &  & Tr.B  & 1.00 & 1.00 & 1.00 \\ 
                    
                            &  & Tr.M  & 1.00 & 1.00 & 1.00 \\ 
                             & & Avg.  & 1.00 & 1.00 & 1.00 \\ \hline
                             \rowcolor{lightblue}
\multirow{5}{*}{PER} & \multirow{5}{*}{1.000}   & IoT.B & 1.00 & 1.00 & 1.00 \\ 
                            &  & IoT.M & 1.00 & 1.00 & 1.00 \\
                            &  & Tr.B  & 1.00 & 1.00 & 1.00 \\ 
                            &  & Tr.M  & 1.00 & 1.00 & 1.00 \\
                            &  & Avg.  & 1.00 & 1.00 & 1.00 \\ \hline
                            \rowcolor{lightblue}
\multirow{5}{*}{MNB} &  \multirow{5}{*}{1.000} & IoT.B & 1.00 & 1.00 & 1.00 \\ 
                            &  & IoT.M & 1.00 & 1.00 & 1.00 \\ 
                            &  & Tr.B  & 1.00 & 1.00 & 1.00 \\ 
                            &  & Tr.M  & 1.00 & 1.00 & 1.00 \\ 
                            &  &  Avg.  & 1.00 & 1.00 & 1.00 \\ \hline
                            \rowcolor{lightblue}
\multirow{5}{*}{BNB}&  \multirow{5}{*}{1.000}  & IoT.B & 1.00 & 1.00 & 1.00 \\ 
                         &  & IoT.M & 1.00 & 1.00 & 1.00 \\
                         &     & Tr.B  & 1.00 & 1.00 & 1.00 \\ 
                         &     & Tr.M  & 1.00 & 1.00 & 1.00 \\ 
                         &     &  Avg.  & 1.00 & 1.00 & 1.00 \\ \hline
\end{tabular}

\begin{tabular}{lllll}
\hline
\multicolumn{5}{c}{\textbf{Without Scaling}}\\ \hline
 \textbf{Acc} & \textbf{Class} & \textbf{Pre} & \textbf{Rec} & \textbf{F1} \\ \hline
   \rowcolor{lightblue}
 \multirow{5}{*}{0.250} & IoT.B & 0.25 & 1.00 & 0.40 \\
                                                        & IoT.M & 0.00 & 0.00 & 0.00 \\
                                                        & Tr.B  & 0.00 & 0.00 & 0.00 \\
                                                         & Tr.M  & 0.00 & 0.00 & 0.00 \\
                                                         & Avg. & 0.06 & 0.25 & 0.10 \\ \hline
    \rowcolor{lightblue}
 \multirow{5}{*}{0.250} & IoT.B & 0.00 & 0.00 & 0.00 \\
                                                        & IoT.M & 0.00 & 0.00 & 0.00 \\
                                                         & Tr.B  & 0.25 & 1.00 & 0.40 \\
                                                         & Tr.M  & 0.00 & 0.00 & 0.00 \\
                                                         & Avg. & 0.06 & 0.25 & 0.10 \\ \hline
      \rowcolor{lightblue}
 \multirow{5}{*}{0.972} & IoT.B & 0.99 & 1.00 & 1.00 \\ 
                                                        & IoT.M & 1.00 & 1.00 & 1.00 \\ 
                                                         & Tr.B  & 1.00 & 0.90 & 0.95 \\ 
                                                         & Tr.M  & 0.91 & 0.99 & 0.95 \\ 
                                                        & Avg. & 0.97 & 0.97 & 0.97 \\ \hline
      \rowcolor{lightblue}
 \multirow{5}{*}{1.000} & IoT.B & 1.00 & 1.00 & 1.00 \\ 
                                                        & IoT.M & 1.00 & 1.00 & 1.00 \\ 
                                                         & Tr.B  & 1.00 & 1.00 & 1.00 \\
                                                        & Tr.M  & 1.00 & 1.00 & 1.00 \\ 
                                                         & Avg. & 1.00 & 1.00 & 1.00 \\ \hline
\end{tabular}
\end{adjustbox}
\end{table}

\begin{figure}[h!]
    \centering
    \includegraphics[width=\columnwidth]{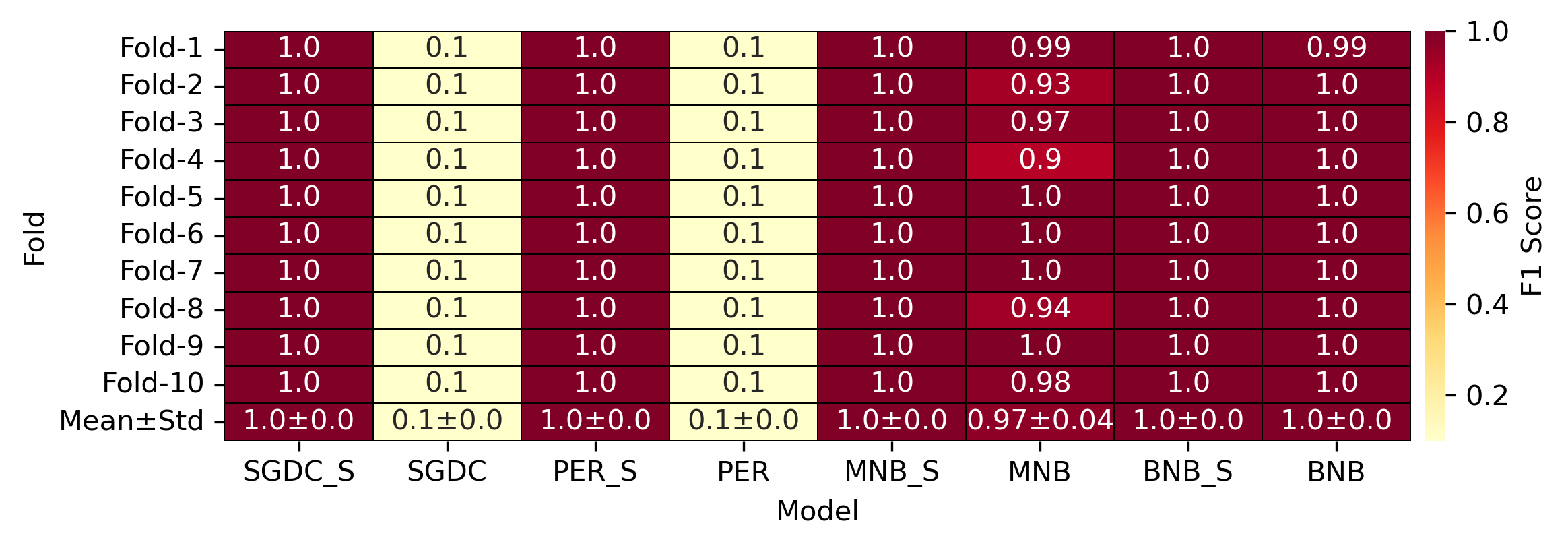}
    \caption{Baseline M1 Models F1 Score using K-Fold Cross Validation }
    \label{tab:classification_report10fold}
\end{figure}

Figure~\ref{tab:classification_report10fold} shows the 10-fold cross-validation performance of M1. Each model was evaluated with and without min-max scaling. Without scaling, SGDC and PER performed poorly (F1 = 0.10 $\pm$ 0.00) due to their sensitivity to feature magnitude, which hinders gradient descent convergence. After scaling, both achieved perfect F1 = 1.00 $\pm$ 0.00. Scaling normalizes feature ranges, such as packet size and traffic rate, ensuring balanced weight updates and improved performance. MNB and BNB also benefited from scaling: MNB improved from 0.97 $\pm$ 0.04 to 1.00 $\pm$ 0.00, while BNB maintained perfect performance in both cases, demonstrating its robustness. Consequently, BNB is selected as the M1 model in our approach.

\begin{figure}[h!]
    \centering
    \includegraphics[width=0.6\linewidth]{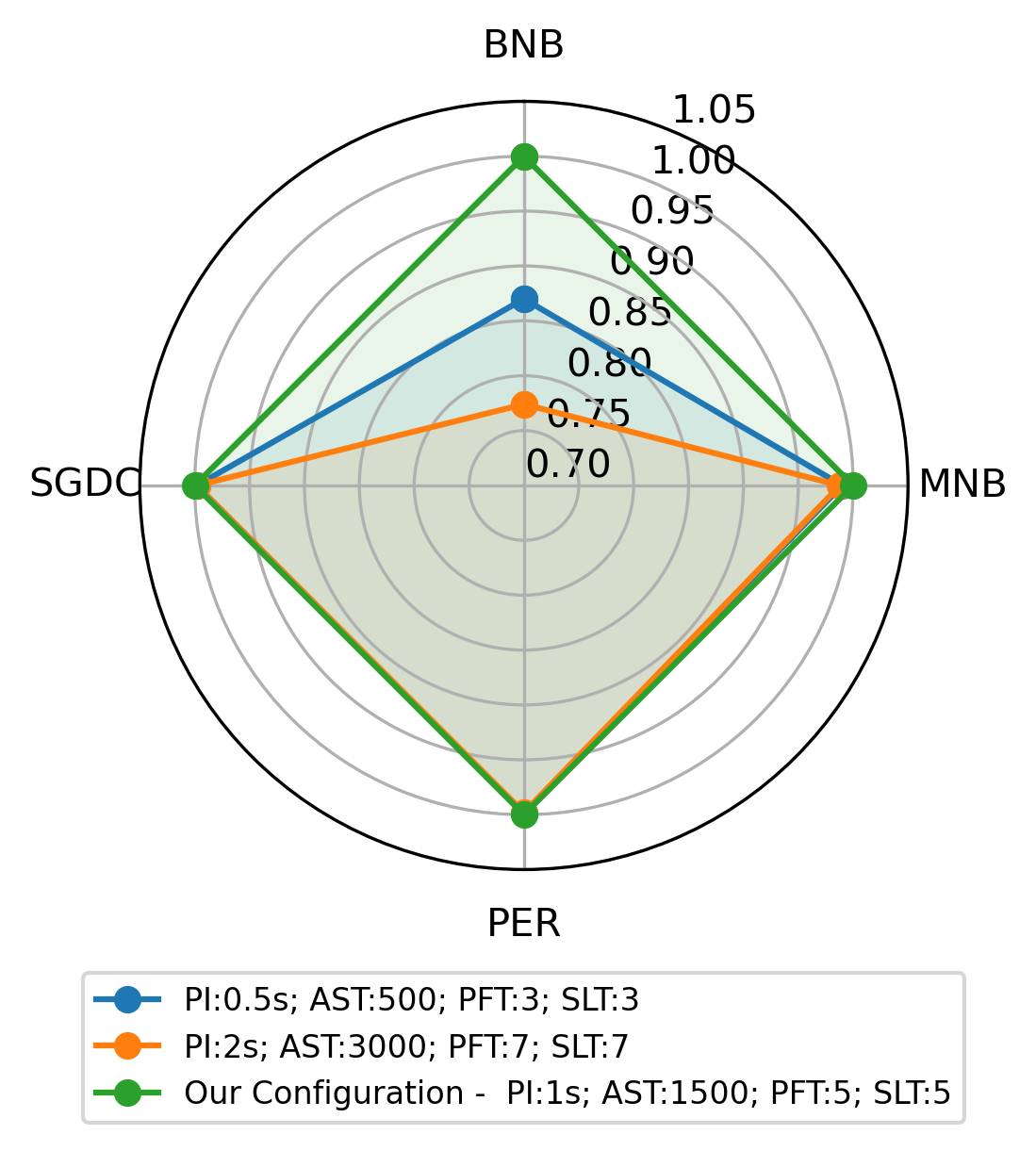}
    \caption{Accuracies Across Three Feature Extraction configurations}
    \label{fig:radar_accuracy}
\end{figure}

To validate the effectiveness of our feature extraction configuration and threshold settings, we conducted experiments using three distinct parameter scenarios. Each scenario varied the processing interval (PI), abnormal size threshold (AST), port frequency threshold (PFT), and short-lived threshold (SLT), all of which directly influence flow-based feature computation from PCAP files. As shown in Figure~\ref{fig:radar_accuracy}, all models achieved perfect accuracy under our balanced configuration, confirming its suitability. PER, SGDC, and MNB exhibited strong robustness across scenarios, while BNB showed greater sensitivity to threshold changes. These results underscore the importance of carefully tuning feature aggregation parameters and validate the reliability and generalizability of our selected configuration.

\subsection{Baseline M2 Models Results}

The performance of M2 models is shown in Table \ref{M2Results}. These results are based on a full dataset consisting of approximately 40 million samples, and we experimented only after data scaling because, in the M1 case, we obtained the best results with data scaling. The performance of all models is significant, with RF achieving a perfect 1.000 accuracy score and outperforming all other evaluation metrics. RF manages many features and determines their importance, which is particularly beneficial in network traffic data that often contains diverse and numerous features. By focusing on the most relevant features, RF enhances predictive accuracy for identifying malicious traffic in our M-En network.
Furthermore, RF scales well with large datasets, as in our case, and handles extensive data efficiently, a common requirement in network traffic analysis. While BNB performed poorly with an accuracy score of 0.8316, all models took considerable time to train but achieved significant results. Our baseline approach for \multiLFramework~ used RF as M2 because of its significant performance.

\begin{table}[h!]
    \centering
    \caption{Results For Baseline M2 Models Using M-En Data}
    \label{M2Results}
    \resizebox{\columnwidth}{!}{
    \begin{tabular}{lclccc}
        \hline
        \textbf{Model} &  \textbf{Acc} &  \textbf{Class} & \textbf{Pre} & \textbf{Rec} & \textbf{F1} \\
        \hline
        \rowcolor{lightblue}
        \multirow{4}{*}{LR} & \multirow{4}{*}{0.999} & IoT.B & 1.00 & 1.00 & 1.00 \\
        && IoT.M & 1.00 & 1.00 & 1.00 \\
        && Tr.B & 1.00 & 1.00 & 1.00 \\
        && Tr.M & 1.00 & 1.00 & 1.00 \\
        &&  Avg & 1.00 & 1.00 & 1.00 \\ \hline
        \rowcolor{lightblue}
        \multirow{4}{*}{SGDC}& \multirow{4}{*}{0.998} & IoT.B & 1.00 & 1.00 & 1.00 \\
        && IoT.M & 1.00 & 1.00 & 1.00 \\
        && Tr.B & 1.00 & 0.99 & 0.99 \\
        && Tr.M & 1.00 & 1.00 & 1.00 \\
        &&  Avg & 1.00 & 1.00 & 1.00 \\ \hline
        \rowcolor{lightblue}
        \multirow{4}{*}{PER}& \multirow{4}{*}{0.999} & IoT.B & 1.00 & 1.00 & 1.00 \\
        && IoT.M & 1.00 & 1.00 & 1.00 \\
        && Tr.B & 1.00 & 1.00 & 1.00 \\
        && Tr.M & 1.00 & 1.00 & 1.00 \\
        &&  Avg & 1.00 & 1.00 & 1.00 \\ \hline
        \rowcolor{lightblue}
        \multirow{4}{*}{MNB}& \multirow{4}{*}{0.994} & IoT.B & 1.00 & 1.00 & 1.00 \\
        && IoT.M & 0.99 & 0.99 & 0.99 \\
        && Tr.B & 0.99 & 0.99 & 0.99 \\
        && Tr.M & 0.99 & 1.00 & 0.99 \\
       &&  Avg & 0.99 & 0.99 & 0.99 \\ \hline
    \end{tabular}
      \begin{tabular}{lclccc}
        \hline
        \textbf{Model} &  \textbf{Acc} &  \textbf{Class} & \textbf{Pre} & \textbf{Rec} & \textbf{F1} \\
        \hline
       \rowcolor{lightblue}
        \multirow{4}{*}{BNB}& \multirow{4}{*}{0.831} & IoT.B & 0.82 & 1.00 & 0.90 \\
        && IoT.M & 0.99 & 0.74 & 0.85 \\
        && Tr.B & 0.80 & 1.00 & 0.89 \\
        && Tr.M & 0.94 & 0.21 & 0.34 \\
       &&  Avg & 0.89 &     0.73  &    0.74\\ \hline 
       \rowcolor{lightblue}
        \multirow{4}{*}{GNB}& \multirow{4}{*}{0.998} & IoT.B & 1.00 & 1.00 & 1.00 \\
        && IoT.M & 1.00 & 1.00 & 1.00 \\
        && Tr.B & 1.00 & 0.99 & 1.00 \\
        && Tr.M & 1.00 & 1.00 & 1.00 \\
       &&  Avg & 1.00 & 1.00 & 1.00 \\ \hline
       \rowcolor{lightblue}
        \multirow{4}{*}{RF}& \multirow{4}{*}{1.000} & IoT.B & 1.00 & 1.00 & 1.00 \\
        && IoT.M & 1.00 & 1.00 & 1.00 \\
        && Tr.B & 1.00 & 1.00 & 1.00 \\
        && Tr.M & 1.00 & 1.00 & 1.00 \\
        &&  Avg & 1.00 & 1.00 & 1.00 \\ \hline
        \rowcolor{lightblue}
        \multirow{4}{*}{ADA}& \multirow{4}{*}{0.998} & IoT.B & 1.00 & 1.00 & 1.00 \\
        && IoT.M & 1.00 & 0.99 & 0.99 \\
        && Tr.B & 0.99 & 0.99 & 0.99 \\
        && Tr.M & 1.00 & 1.00 & 1.00 \\
       &&  Avg & 1.00 & 1.00 & 1.00 \\ \hline
    \end{tabular}}
    \label{tab:classification_report}
\end{table}


The performance metrics for the M2 models in terms of \emph{correct prediction} (CP), \emph{wrong prediction} (WP), and error rate are presented in Table \ref{tab:model_performanceCM}. RF model demonstrated the most significant performance with the highest CP of 10,146,052 and the lowest error rate of 0.0000, which justifies its selection as M2 in our proposed approach. LR and PER models both performed admirably, with error rates of 0.0005, indicating their reliability in malicious traffic detection. The GNB and ADA models also showed strong performance with error rates of 0.0016 and 0.0013, respectively. On the other hand, the BNB model had a significantly higher error rate of 0.1684, indicating its limited effectiveness in this context. The MNB model also showed a relatively higher error rate of 0.0057 compared to other models. These results highlight the robustness of the RF model and the varying degrees of effectiveness among different ML models in detecting malicious traffic.

\begin{table}[h!]
    \centering
        \caption{Baseline M2 Models Performance Metrics}
        \resizebox{0.5\columnwidth}{!}{
    \begin{tabular}{lllll}
        \hline
        \textbf{Model} & \textbf{CP} & \textbf{WP} & \textbf{Error Rate} \\
        \hline
            \rowcolor{lightblue}
        LR & 10,140,676 & 5,395 &  0.0005 \\
        SGDC & 10,128,182 & 17,889 &  0.0018 \\
            \rowcolor{lightblue}
        PER & 10,140,625 & 5,446 &  0.0005 \\
        MNB & 10,088,644 & 57,427 &  0.0057 \\
            \rowcolor{lightblue}
        BNB & 8,437,084 & 1,708,987 &  0.1684 \\
        GNB & 10,130,116 & 15955 &  0.0016 \\
            \rowcolor{lightblue}
        RF & 10,146,052 & 19 &  0.0000 \\
        ADA & 10,132,709 & 13362 &  0.0013 \\
        \hline
    \end{tabular}}
    \label{tab:model_performanceCM}
\end{table}

Table \ref{tab:model_performanceTime} shows the computational time required for the training and testing of baseline models used at the M1 and M2 levels, measured in seconds. The M2 models were trained on a larger dataset and thus exhibited higher computational times compared to the M1 models. In the M2 category, LR and ADA took significantly longer, with times of 13,815.6875 seconds and 14,106.234375 seconds, respectively. RF also required considerable time, 12,841.046875 seconds, reflecting its complex ensemble nature. Comparatively, models like SGDC, PER, MNB, and BNB showed much lower computational times, highlighting their relative efficiency for the same tasks. However, RF is significant in terms of accuracy and also average computational cost, so we chose it for the proposed approach from M2 and BNB from M1.

\begin{table}[h!]
    \centering
        \caption{Baseline M1 \& M2 Models Computational Time}
        \resizebox{0.35\columnwidth}{!}{
    \begin{tabular}{lll}
        \hline
        \textbf{Model} & \textbf{M1} & \textbf{M2} \\
        \hline
            \rowcolor{lightblue}
        LR & - &  138.68 \\
        SGDC & 1.17  & 325.17  \\
            \rowcolor{lightblue}
        PER & 0.93  & 253.45  \\
        MNB & 1.12 & 113.29  \\
            \rowcolor{lightblue}
        BNB & 1.17  & 124.85 \\
        GNB & - & 71.0   \\
            \rowcolor{lightblue}
        RF & - &  12841.04 \\
        ADA & - &  14106.23 \\
        \hline
    \end{tabular}}
    \label{tab:model_performanceTime}
\end{table}

\subsection{Results in Real-Time Environment}
\label{resultstime}
\begin{figure}[h!]
    \centering
    \includegraphics[width=\columnwidth]{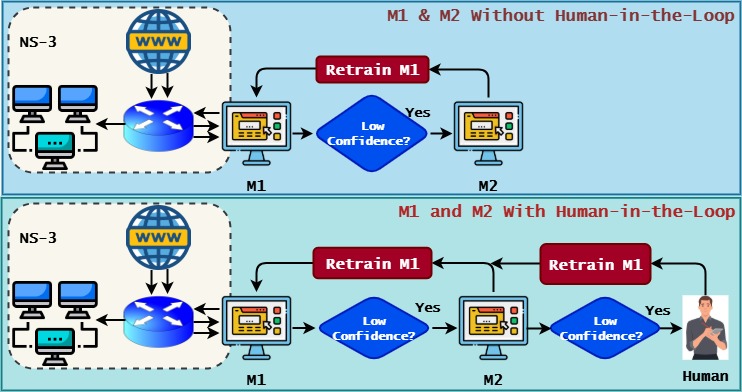}
    \caption{\textit{Scenario 4 \& 5} Visualization For Easy Understanding }
    \label{figMLF}
\end{figure}

In this section, we present the results of \multiLFramework~in real-time scenarios. For real-time testing, we used NS-3 simulation to generate sample data. The data was collected centrally and then passed to \multiLFramework. We conducted experiments under different scenarios to evaluate the framework's performance. We collected packets over specific time windows and then passed the batch of packets to the framework to determine the accuracy. This process was repeated over several iterations, and the average scores were reported.
We select the best models from online testing, evaluate their performance, and compare them across different scenarios.
\begin{enumerate*}[label=(\textit{\roman*})]
\item \textit{\textbf{Scenario 1:}} M1 is used for attack detection without continuous learning capabilities and evaluated with benign and malicious traffic. 

\item \textit{\textbf{Scenario 2:}} M1 is used for attack detection with continuous learning capabilities. 

\item \textit{\textbf{Scenario 3:}} M2 is used independently for testing. 

\item \textit{S\textbf{cenario 4:}} M1 and M2 are deployed together under \multiLFramework~without human involvement. 

\item \textit{\textbf{Scenario 5:}} M1 and M2 are deployed under \multiLFramework~with human involvement. The implementation of Scenarios 4 and 5 in \multiLFramework~is illustrated in Figure \ref{figMLF}.
\end{enumerate*}

Table~\ref{tabM1RT} provides an evaluation of \textit{\textbf{Scenario 1}}, where the M1 model is used without continuous learning for attack detection. KNN demonstrates the highest accuracy of 0.688 but at the cost of substantial resource usage, including a large MS 90,236 KB, high memory consumption 187.741 MB, and significant CPU utilization 58.34\%. In contrast, models like LR exhibit a much smaller size of 2 KB, a lower MU of 122.3 MB, and a minimal CPU impact of 45.61\% with a faster prediction time of 0.0007 seconds, with reduced accuracy of 0.536. These results highlight the trade-off between model complexity and computational efficiency, making KNN suitable for accuracy-critical scenarios, while lighter models, such as LR, are better suited for resource-constrained environments.

\begin{table}[h!]
\centering
\caption{\textbf{Scenario 1:} Baseline M1 Without Continuous Learning }
\label{tabM1RT}
\resizebox{0.85\columnwidth}{!}{
\begin{tabular}{llllll}
\hline
\textbf{Model} & \textbf{ADA} & \textbf{GNB} & \textbf{KNN} & \textbf{LR} & \textbf{RF} \\ \hline
    \rowcolor{lightblue}
MS (KB) & 57 & 4 & 90236 & 2 & 103 \\ 
Accuracy  & 0.32 & 0.422 & 0.688 & 0.536 & 0.661 \\ 
    \rowcolor{lightblue}
PT (S) & 0.071 & 0.002 & 0.723 & 0.0007 & 0.032\\
CPU  (\%) & 52.91 & 49.12 & 58.34 & 45.61 & 51.29 \\ 
    \rowcolor{lightblue}
MU (MB) & 122.467 & 121.571 & 187.741 & 122.3 & 112.81 \\ \hline
\end{tabular}}
\end{table}

\textit{\textbf{Scenario 2}} evaluates the M1 model with continuous learning capabilities, showcasing improvements in both efficiency and performance as shown in Table \ref{tabM1ILRT}. Among the models tested, PER achieves the highest accuracy of 0.704 while maintaining a minimal MS of 3 KB and a low prediction time of 0.0007 seconds. These results demonstrate the model's ability to quickly adapt to new data without a significant increase in computational costs. Other models, such as MNB and SGDC, also display competitive accuracies of 0.594 and 0.581, respectively, while maintaining a low memory footprint of around 117-120 MB. These results highlight the effectiveness of continuous learning in maintaining high accuracy while optimizing resource usage for real-time scenarios, as it achieves an accuracy of 0.704 compared to \textit{\textbf{Scenario 1}}, which has an accuracy of 0.688.

\begin{table}[h!]
\centering
\caption{\textbf{Scenario 2:} Baseline M1 With Continuous Learning }
\resizebox{0.8\columnwidth}{!}{
\label{tabM1ILRT}
\begin{tabular}{lrrrr}
\hline
\textbf{Model} & \textbf{BNB} & \textbf{MNB} & \textbf{PER} & \textbf{SGDC} \\ \hline
    \rowcolor{lightblue}
MS (KB) & 4 & 4 & 3 & 3 \\ 
Accuracy  & 0.575 & 0.594 & 0.704 & 0.581 \\ 
    \rowcolor{lightblue}
PT (S) & 0.001 & 0.001 & 0.0007 & 0.001 \\ 
CPU (\%) & 49.69 & 50.01 & 55.52 & 51.74 \\ 
    \rowcolor{lightblue}
MU (MB) & 110.248 & 117.812 & 124.405 & 120.932 \\ \hline
\end{tabular}}
\end{table}


In \textit{\textbf{Scenario 3}}, the performance of M2 is evaluated without continuous learning across multiple machine learning models, as shown in Table~\ref{tab1Sc3}. The results show notable variation in accuracy, prediction time, CPU utilization, and memory usage. MLP achieves the highest accuracy of 0.885 but requires high CPU utilization of 57.14\% and memory usage of 140.12 MB. LR offers balanced performance with an accuracy of 0.818 and low computational cost, making it suitable for resource-constrained environments. RF attains moderate accuracy of 0.738 but consumes the most resources, with an MS of 33,009 KB and MU of 267.68 MB, highlighting a trade-off between accuracy and efficiency. BNB and GNB achieve accuracies of 0.725 and 0.721 respectively, maintaining low prediction times and memory use, though their high CPU utilization limits real-time applicability. Compared to \textit{\textbf{Scenario 2}}, where continuous learning was employed, \textit{\textbf{Scenario 3}} shows that M2 can achieve higher accuracy with some models, such as MLP, but at the cost of increased computational overhead. This emphasizes the importance of continuous learning for optimizing resource usage and ensuring real-time performance. Consequently, M1 with continuous learning is applied at the first level, and the best M2 model is deployed at the second level in the \multiLFramework.

\begin{table}[h!]
\centering
\caption{\textbf{Scenario 3:} Baseline M2 Performance Comparison }
\label{tab1Sc3}
\resizebox{0.85\columnwidth}{!}{
\begin{tabular}{lrrrrr}
\hline
\textbf{Model} &MS (KB) & Acc  & PT (S) & CPU (\%) & MU (MB) \\ \hline
    \rowcolor{lightblue}
ADA & 57 & 0.723 & 0.085 & 39.91 & 96.179 \\ 
BNB & 4 & 0.725 & 0.001 & 60.47 & 129.028 \\ 

    \rowcolor{lightblue}
GNB & 4 & 0.721 & 0.005 & 56.28 & 137.565 \\ 
LR & 2 & 0.818 & 0.001 & 48.16 & 123.635 \\ 
    \rowcolor{lightblue}
MLP & 106 & 0.885 & 0.016 & 57.14 & 140.119 \\ 
MNB & 4 & 0.792 & 0.0009 & 53.57 & 133.187 \\ 
    \rowcolor{lightblue}
PER & 3 & 0.737 & 0.0009 & 61.79 & 151.869 \\ 
RF & 33009 & 0.738 & 0.0313 & 55.18 & 267.679 \\ 
    \rowcolor{lightblue}
SGDC & 3 & 0.830 & 0.002 & 46.00 & 137.654 \\ \hline
\end{tabular}
}
\end{table}

Table \ref{tab4MultiLF} shows the performance comparison of four models, BNB, MNB, PER, and SGDC, under \textit{\textbf{Scenario 4}}, where both M1 and M2 models are deployed in \multiLFramework~without human involvement. The first set of results evaluates the models with MLP as the M2 because of its significant performance. The PER model achieves the highest accuracy at 0.999, but at the cost of significantly higher CPU utilization, 23.22\%, and MU 2.623 MB, indicating its computational intensity. In contrast, BNB maintains high accuracy at 0.975 with a much lower CPU usage of 2.582\% and MU 1.513 MB, suggesting a good trade-off between performance and resource efficiency.

The second set of results involves the same models but with RF as the M2. The accuracy of all models drops compared to the MLP scenario, with the highest accuracy of 0.983, achieved by the PER model. However, the computational overhead is significantly reduced, as evidenced by lower CPU utilization, 3.221\%, and MU 0.956 MB. These results suggest that using RF as the secondary model is less resource-intensive, but it comes at the cost of reduced prediction performance. The results highlight the performance, efficiency trade-offs between using MLP and RF as M2 models in \multiLFramework. The PER model consistently achieves high accuracy but at a higher computational cost, while BNB and other models offer a balanced approach between accuracy and resource usage.

\begin{table}[h!]
\centering
\caption{\textbf{Scenario 4:} Baseline M1 and M2 Under \multiLFramework~Without Human-in-the-Loop}
\resizebox{0.85\columnwidth}{!}{
\begin{tabular}{lcccc}
\hline
\multicolumn{5}{c}{\textbf{MLP →  M2 \& Without Human-in-the-Loop}}\\\hline
Matrix & BNB &	MNB &	PER &	SGDC \\ \hline
    \rowcolor{lightblue}
Accuracy & 0.975 & 0.331 & 0.999 & 0.371 \\ 
PT (S) & 0.332 & 0.270 & 1.259 & 0.144 \\ 
    \rowcolor{lightblue}
CPU  (\%) & 2.582 & 0.027 & 23.22 & 0.034 \\ 
MU (MB) & 1.513 & 0.041 & 2.623 & 0.030 \\ \hline
\multicolumn{5}{c}{\textbf{RF → M2 \& Without Human-in-the-Loop}}\\\hline
Matrix & BNB &	MNB &	PER &	SGDC \\ \hline
    \rowcolor{lightblue}
Accuracy  & 0.941 & 0.330 & 0.983 & 0.186 \\ 
PT (S) & 0.216 & 0.270 & 0.278 & 0.164 \\ 
    \rowcolor{lightblue}
CPU (\%) & 1.240 & 0.023 & 3.221 & 0.042 \\ 
MU (MB) & 1.338 & 0.017 & 0.956 & 0.001 \\ \hline
\end{tabular}}
\label{tab4MultiLF}
\end{table}

In \textit{\textbf{Scenario 5}}, M1 and M2 models were evaluated under the \multiLFramework~with human involvement. We used BNB, MNB, PER, and SGDC as M1  with MLP and RF as the M2, same as \textit{\textbf{Scenario 4}}. This scenario demonstrated the impact of incorporating human oversight on model performance. When MLP was used as the M2, the PER achieved the highest accuracy of 0.999, but it required significant CPU utilization of 17.84\% and MU of 35.48 MB. The prediction time of 0.866 seconds indicates that while human involvement boosts the model's accuracy, it also increases computational costs. In contrast, the BNB model showed good performance with an accuracy of 0.979, but its MU of 56.53 MB remained notably high compared to other models. When RF was used as the M2, PER again demonstrated high accuracy of 0.983, but with a reduced CPU utilization of 10.05\% and a lower MU of 3.632 MB. Prediction time was also reduced to 0.434 seconds, indicating that RF as the M2 results in a more resource-efficient configuration while maintaining high accuracy. BNB's performance with RF was similar to its performance with MLP in terms of accuracy, but it achieved lower CPU utilization, 2.564\%, and higher MU of 72.69 MB. Comparing \textit{\textbf{Scenario 5}} to \textit{\textbf{Scenario 4}} under \multiLFramework, we observe that human intervention slightly increases computational resource usage but improves model accuracy, demonstrating the value of human oversight. Furthermore, \textit{\textbf{Scenario 5}} outperformed \textit{\textbf{Scenario 1}} and \textit{Scenario 2} (M1 configurations) in accuracy across most models, validating the robustness of the proposed human-involved framework. Finally, compared to \textit{\textbf{Scenario 3}} (M2 without continuous learning), \textit{\textbf{Scenario 5}} exhibited a more balanced trade-off between accuracy and computational efficiency, especially when using RF as the M2, highlighting the advantage of human intervention in refining the model's performance.

\begin{table}[h!]
\centering
\caption{\textbf{Scenario 5:} Baseline M1 and M2 Under \multiLFramework~With Human-in-the-Loop}
\resizebox{0.85\columnwidth}{!}{
\begin{tabular}{ccccc}
\hline
\multicolumn{5}{c}{MLP → M2 \& Human-in-the-Loop}\\\hline
Matrix & BNB &	MNB &	PER &	SGDC \\ \hline
    \rowcolor{lightblue}
Accuracy & 0.979 & 0.331 & 0.999 & 0.371 \\ 
PT (S)  & 0.374 & 0.270 & 0.866 & 0.174 \\ 
    \rowcolor{lightblue}
CPU  (\%) & 3.158 & 0.022 & 17.84 & 0.035 \\ 
MU (MB) & 56.53 & 0.009 & 35.48 & 0.018 \\ \hline
\multicolumn{5}{c}{RF → M2 \& Human-in-the-Loop}\\\hline
    \rowcolor{lightblue}
Accuracy & 0.975 & 0.331 & 0.983 & 0.384 \\ 
PT (S)  & 0.331 & 0.270 & 0.434 & 0.178 \\ 
    \rowcolor{lightblue}
CPU  (\%) & 2.564 & 0.019 & 10.05 & 0.102 \\ 
MU (MB) & 72.69 & 0.020 & 3.632 & 0.020 \\ \hline
\end{tabular}}
\end{table}

\textbf{Human Intervention Feasibility:} In our experimental setup, approximately 111,644 samples are processed across 15 batches. Human intervention is required for only 109 samples, resulting in a human effort percentage of just 0.0976\%. This behavior shows that the approach is highly feasible for real-time deployment, as the burden on human input remains minimal, as shown in Table \ref{tab:hinter}.

\begin{table}[h!]
    \centering
\caption{Human Workload Under \multiLFramework~ Using Baseline M1 \& M2 }
\begin{threeparttable}
    \begin{tabular}{lc}\hline
    \textbf{Metric} & \textbf{M1 → M2 → Human}\\\hline
    \rowcolor{lightblue}
     M1 TRP  & 111,644 \\
     M1 CPP   & 111,153  \\
     \rowcolor{lightblue}
     M2 TRP  & 491\\
     M2 CPP   & 382\\
     \rowcolor{lightblue}
     HIR     & 109 \\
     HE      & 0.0976\%\\
      \hline
    \end{tabular}
\begin{tablenotes}
\scriptsize
\item
\textbf{TRP}-- Total Received Packets; 
\textbf{CPP}-- Correctly Predicted Packets;
\textbf{HIR}-- Human Intervention Required;
\textbf{HE}-- Human Efforts
\end{tablenotes}
\end{threeparttable}
    \label{tab:hinter}
\end{table}

\begin{figure}[h!]
\centering
    \begin{subfigure}[b]{1.7in}
    \centering
    \includegraphics[width=1.7in]{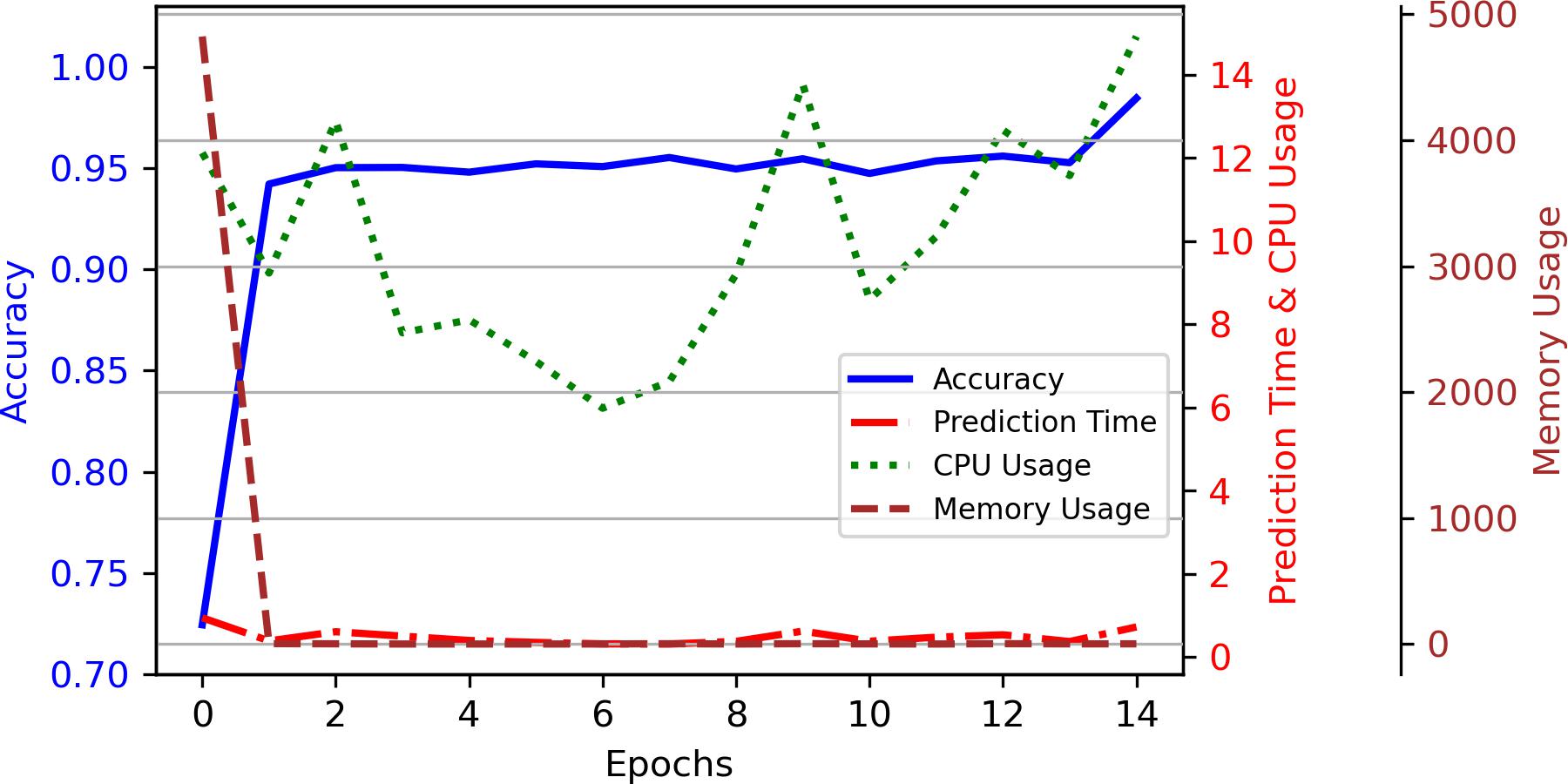}
   \caption{ M1 (BNB) \& M2 (RF) without Human-in-the-Loop}
    \label{fig}
    \end{subfigure}
    \begin{subfigure}[b]{1.7in}
    \centering
    \includegraphics[width=1.7in]{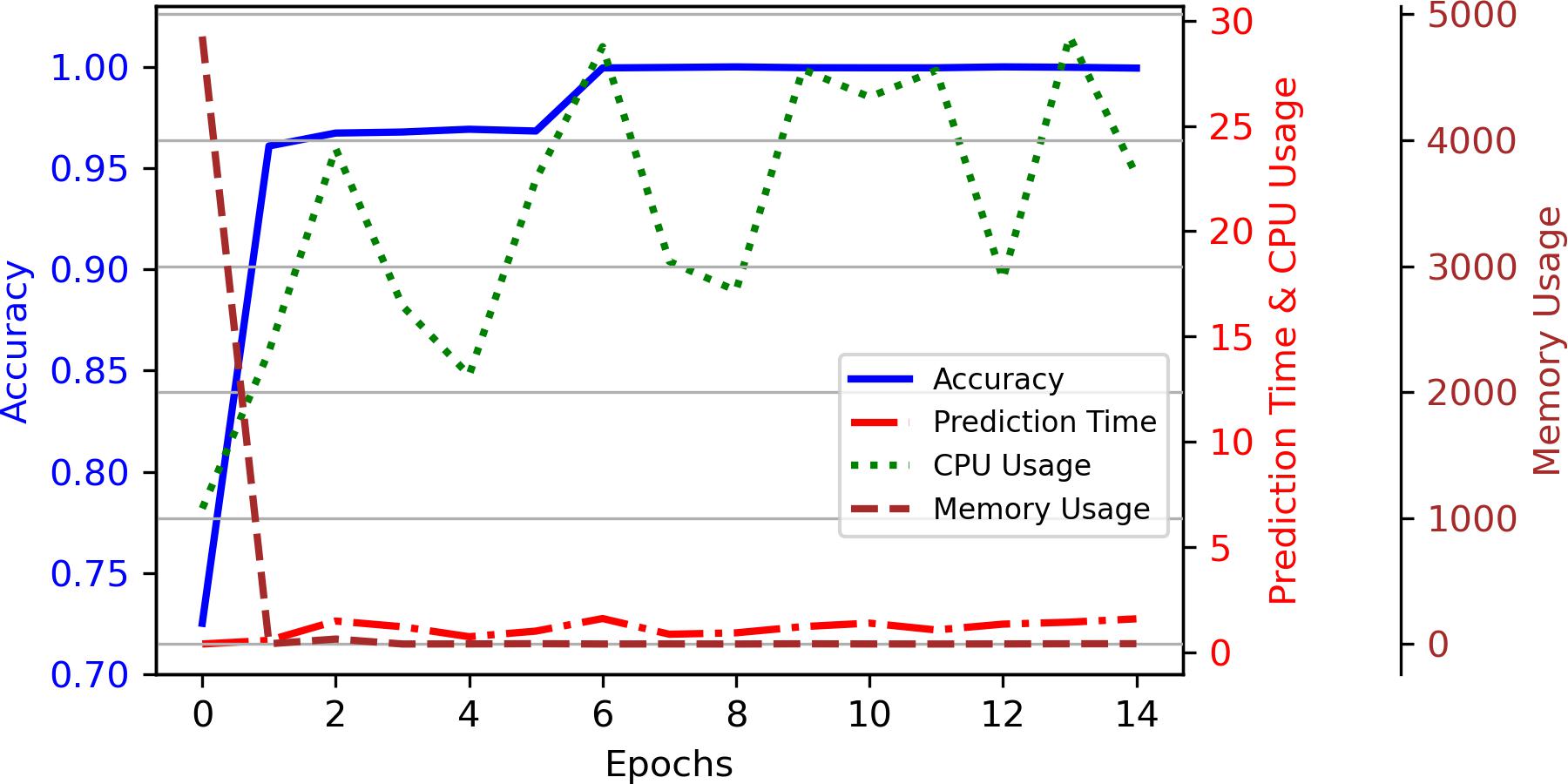}
\caption{ M1 (BNB) \& M2 (MLP) without Human-in-the-Loop}    \label{fig1imp}
    \end{subfigure}

    \begin{subfigure}[b]{1.7in}
    \centering
    \includegraphics[width=1.7in]{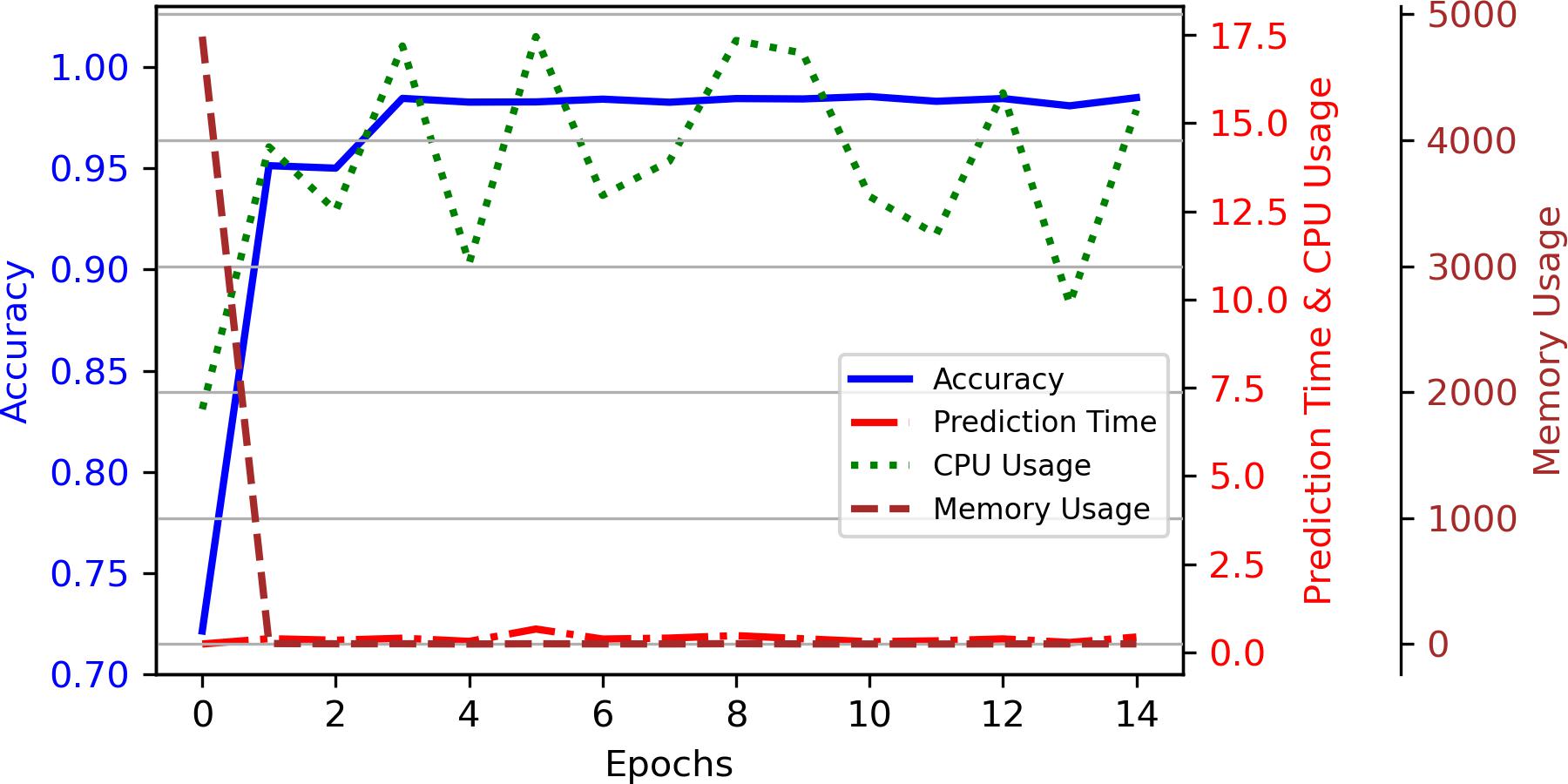}
\caption{M1 (BNB) \& M2 (RF) with Human-in-the-Loop}    \label{fig2imp}
    \end{subfigure}
    \begin{subfigure}[b]{1.7in}
    \centering
    \includegraphics[width=1.7in]{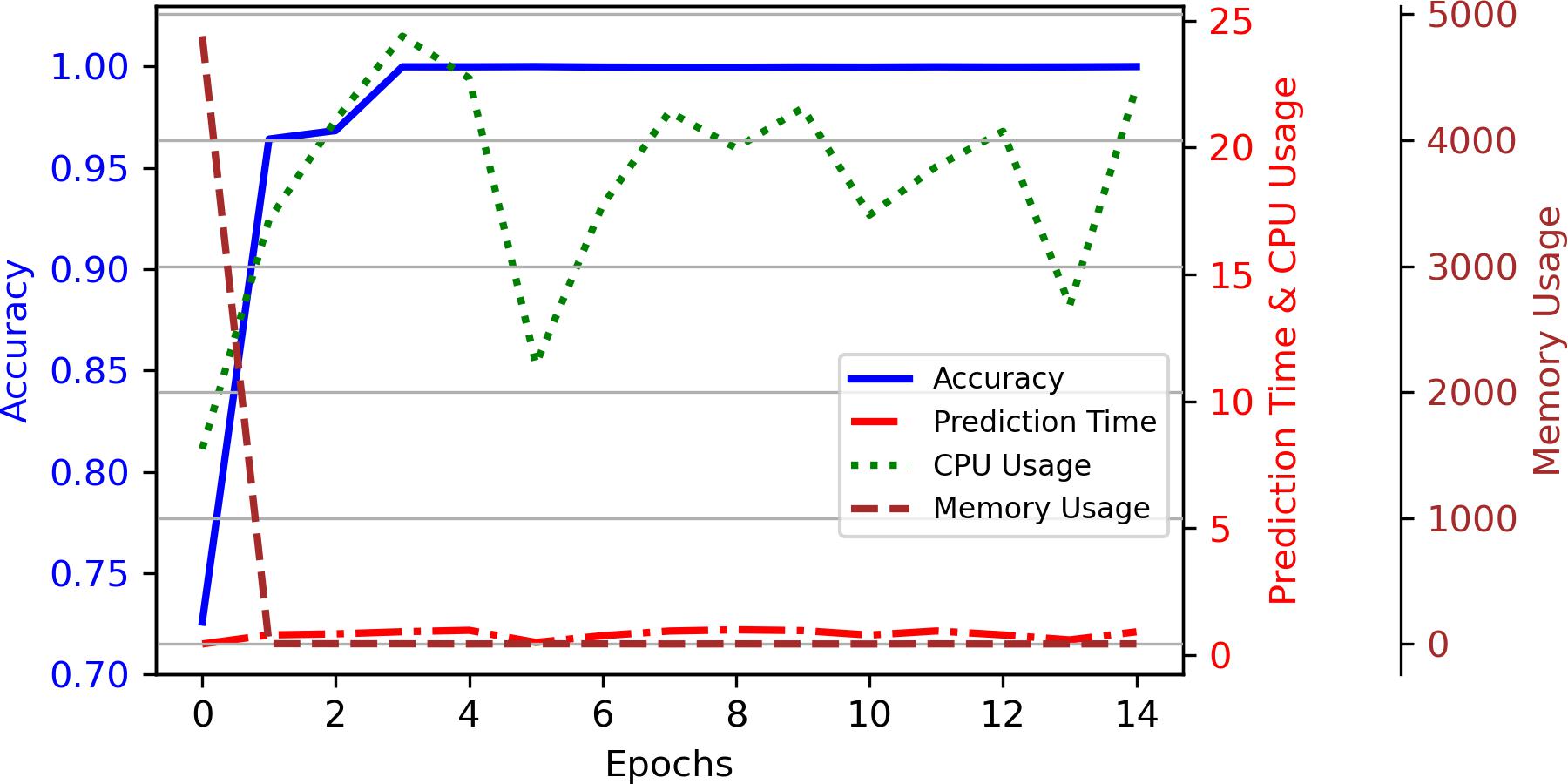}
\caption{M1 (BNB) \& M2 (MLP) with Human-in-the-Loop}    \label{fig3}
    \end{subfigure}

      \begin{subfigure}[b]{1.7in}
    \centering
    \includegraphics[width=1.7in]{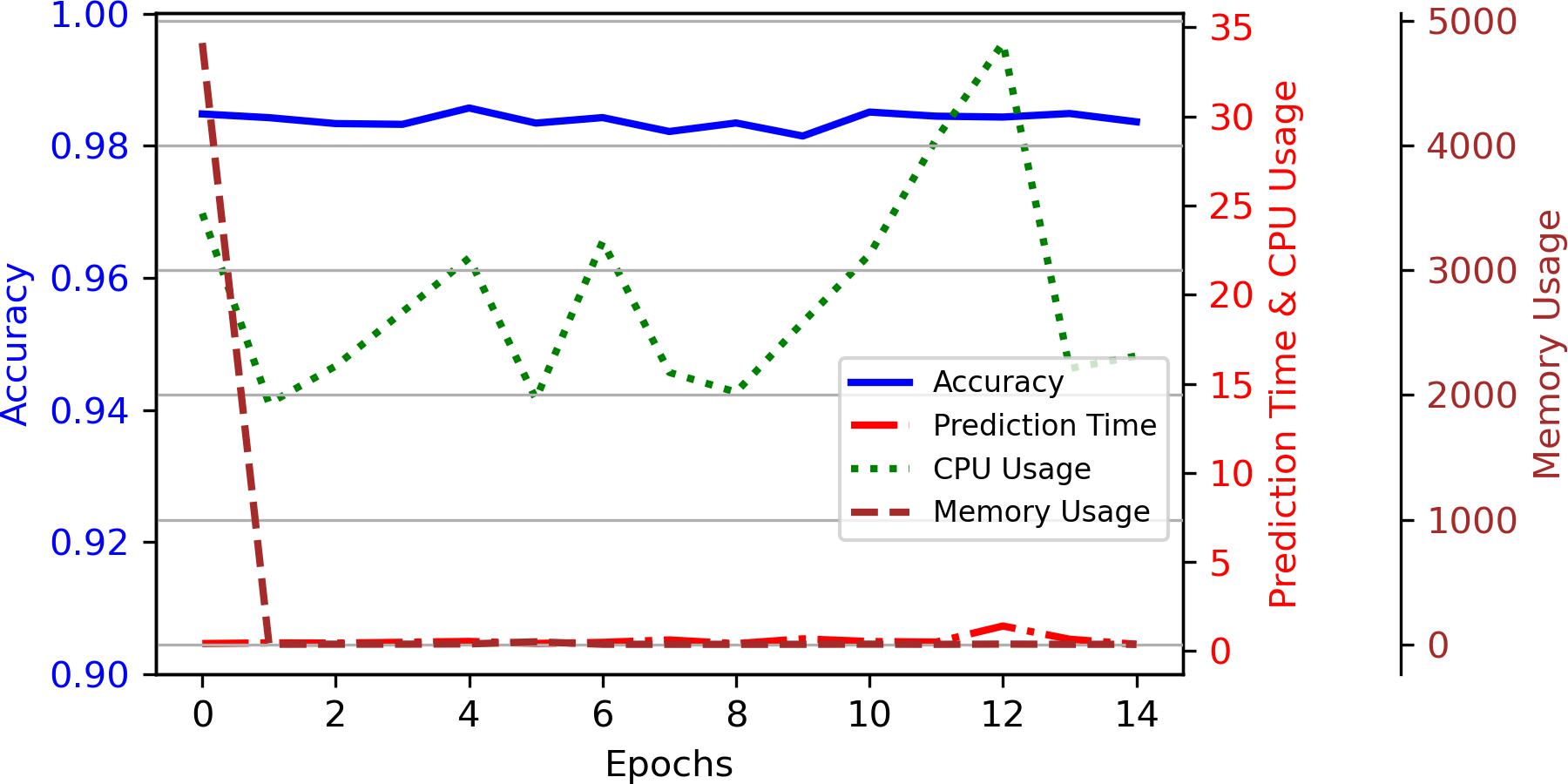}
\caption{M1 (PER) \& M2 (RF) without Human-in-the-Loop}    \label{fig4}
    \end{subfigure}
    \begin{subfigure}[b]{1.7in}
    \centering
    \includegraphics[width=1.7in]{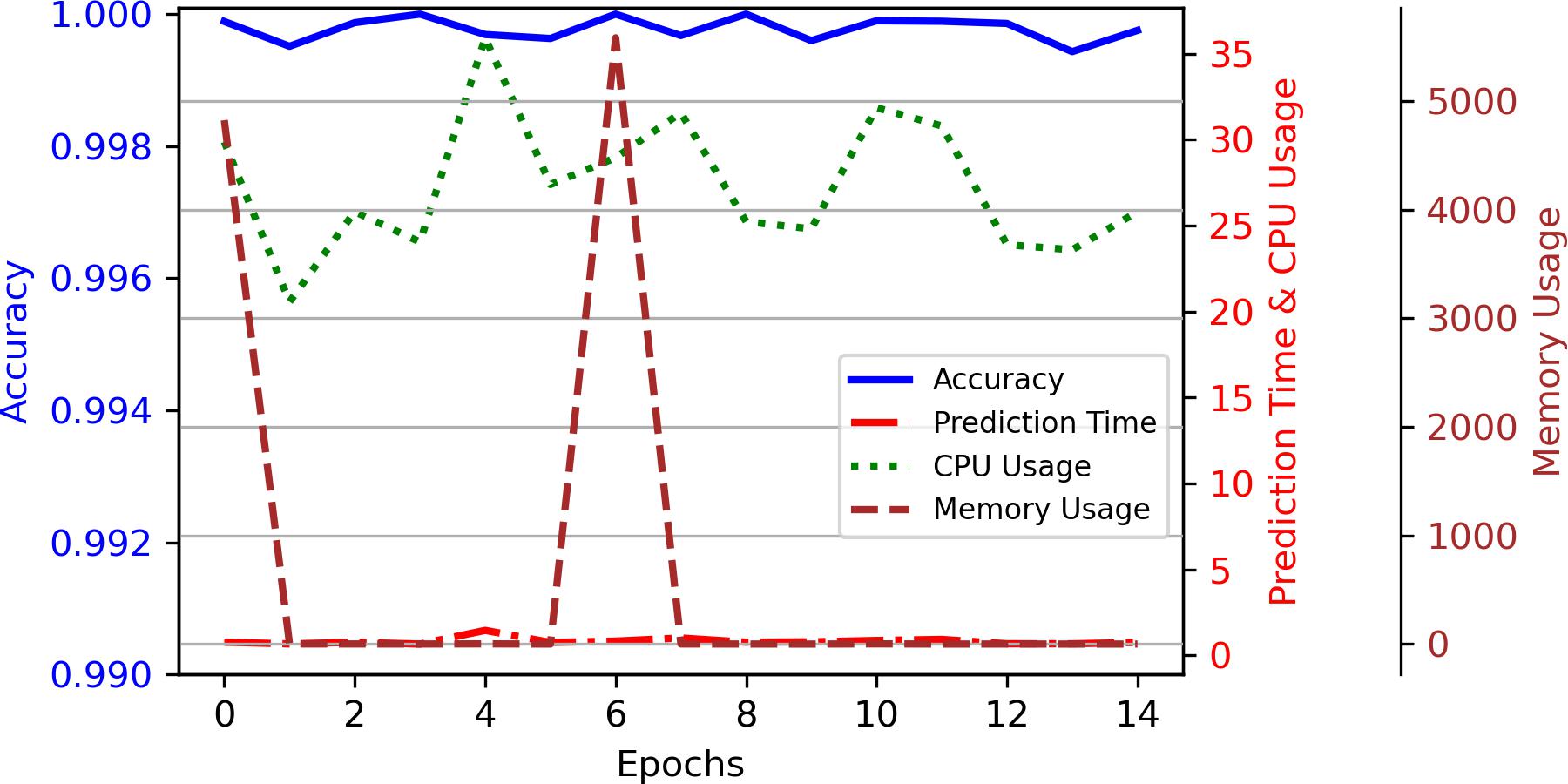}
\caption{M1 (PER) \& M2 (MLP) without Human-in-the-Loop}     \label{fig5}
    \end{subfigure}

      \begin{subfigure}[b]{1.7in}
    \centering
    \includegraphics[width=1.7in]{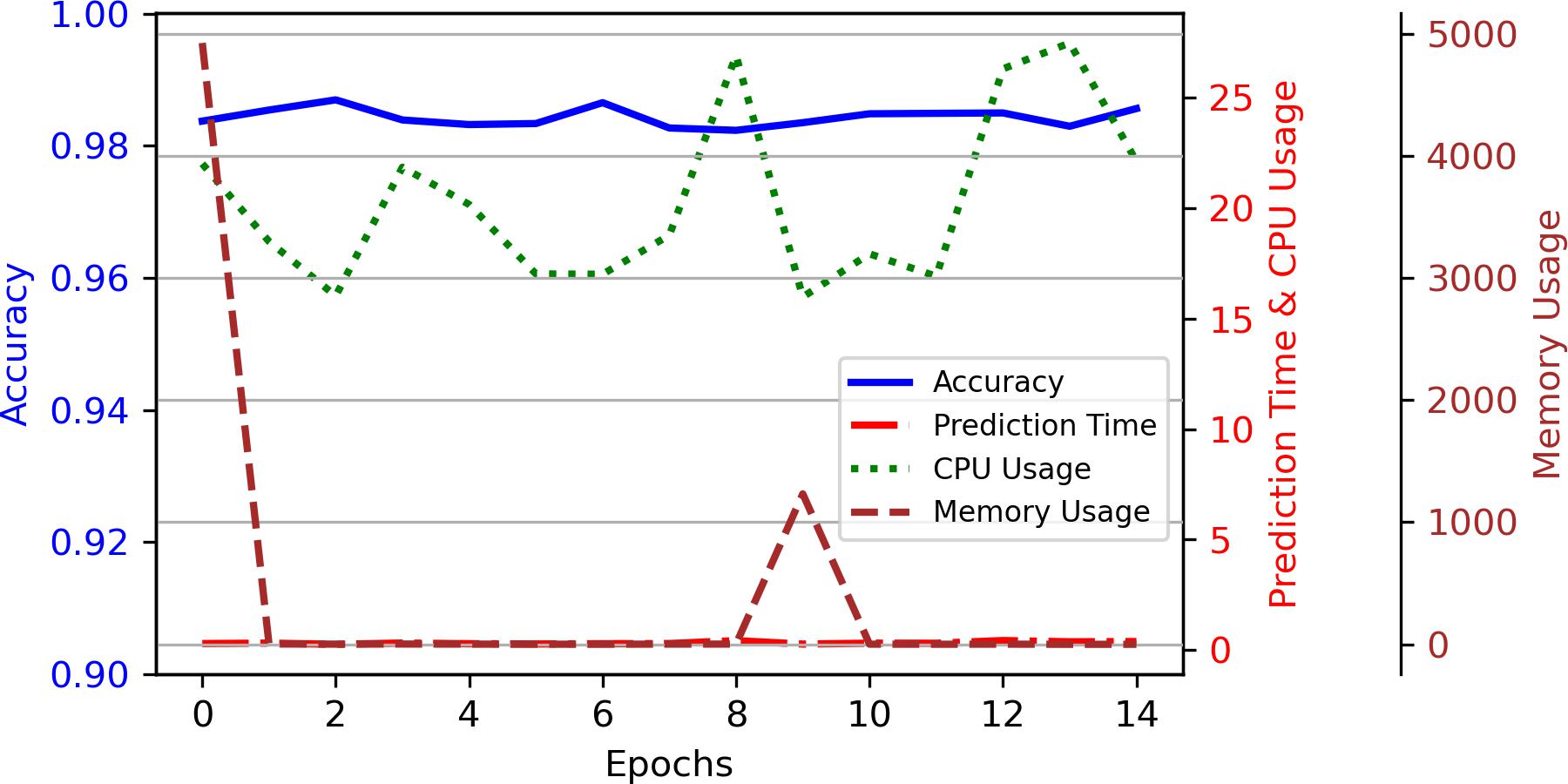}
\caption{M1 (PER) \& M2 (RF) with Human-in-the-Loop}     \label{fig6}
    \end{subfigure}
    \begin{subfigure}[b]{1.7in}
    \centering
    \includegraphics[width=1.7in]{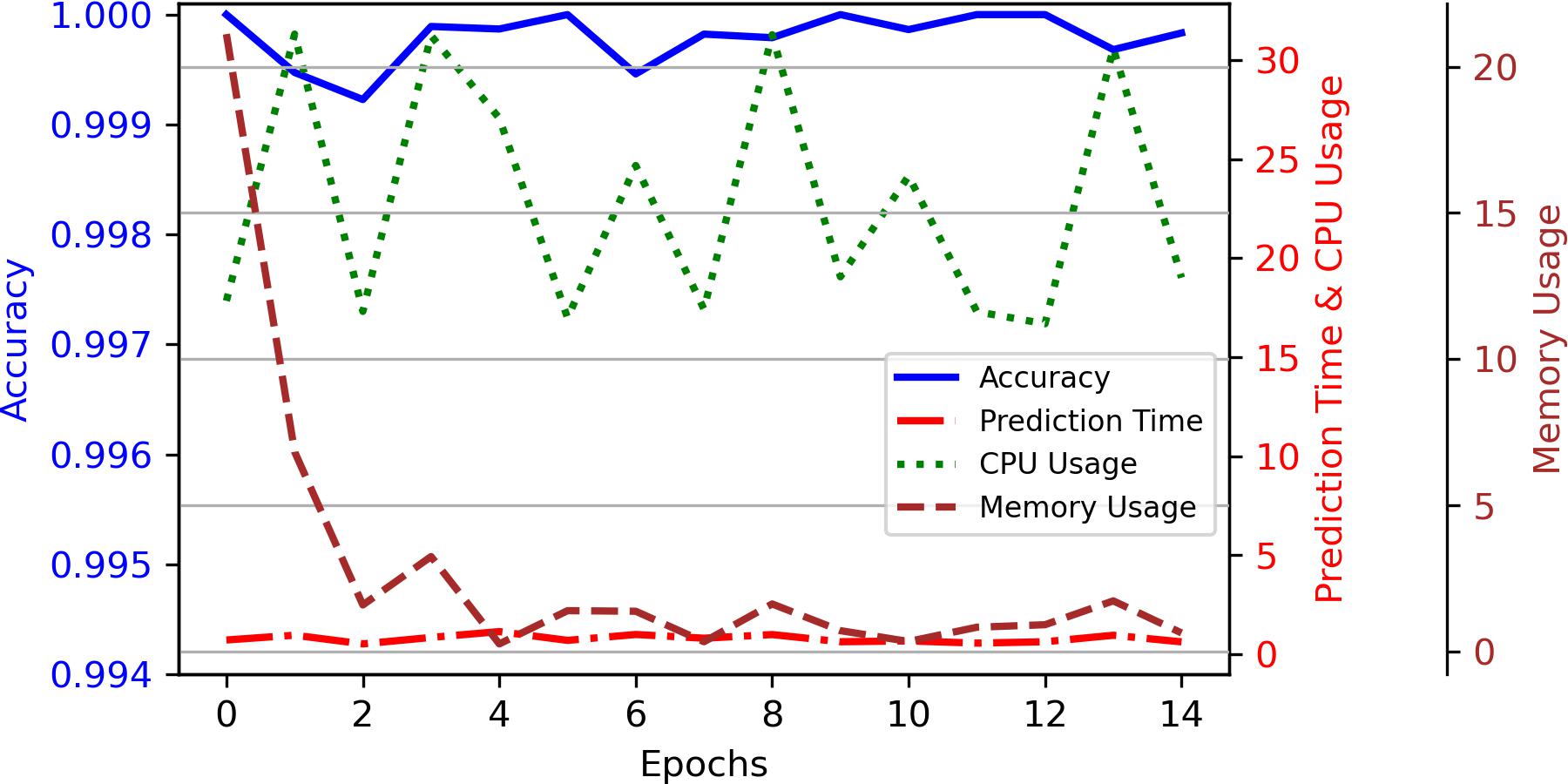}
\caption{ M1 (PER) \& M2 (MLP) with Human-in-the-Loop}     \label{fig7}
    \end{subfigure}
\caption{Per-Batch Results of Best Models For \textit{Scenarios 4} and \textit{5}}
\label{import}
\end{figure}

Figure \ref{import} shows the different iteration scores for Scenarios 4 and 5. It is evident that model accuracy improves with continuous learning as more data is provided. However, the prediction time remains consistent over the iterations.

\subsection{Proposed DL Models Results Under \multiLFramework~Framework}

In addition to traditional machine learning models, we implemented several deep learning architectures in our proposed \multiLFramework~ to enhance malicious traffic detection performance. These include T1, T2, L1, and L2, where T1 and L1 represent lightweight models (used as M1) and T2 and L2 represent more complex variants (used as M2). Deep learning models showed significant performance improvements compared to traditional baselines, largely due to their ability to capture long-term dependencies and complex patterns in the data. The lightweight variants were specifically designed to maximize throughput while maintaining high accuracy. Table~\ref{tabDLbased} shows DL models' performance achieved nearly identical score values, exceeding 0.999, indicating robust and reliable performance. The lightweight models (T1 and L1) show higher throughput, especially T1, which achieved a throughput of 66,130 packets/sec, significantly higher than L1 (6,157 packets/sec) and the complex models T2 and L2. Notably, T1 had a slightly higher error rate of 0.0651, compared to L1, which is 0.0612, but the difference is minimal and offset by its efficiency advantage.

\begin{table}[h!]
    \centering
    \caption{Results Using Proposed DL-based Models }
    \resizebox{\columnwidth}{!}{
    \begin{tabular}{ccccccc}\hline
       \textbf{Model}  &  \textbf{Acc} & \textbf{Pre} & \textbf{Rec} & \textbf{F1} & \textbf{Error Rate} & \textbf{Throughput} \\\hline
    \rowcolor{lightblue}
       T1 &  0.9993 & 0.9994  & 0.9993 & 0.9993 & 0.0651 & 66130 \\
       T2  & 0.9994 & 0.9994 & 0.9994 & 0.9994 & 0.0612 & 65595 \\
    \rowcolor{lightblue}
       L1 & 0.9994 & 0.9994 & 0.9994 & 0.9994 & 0.0612 & 6157 \\
       L2 & 0.9994 & 0.9995 & 0.9994 & 0.9994 & 0.0611 & 4263 \\
       \hline
    \end{tabular}}
    \label{tabDLbased}
\end{table}

Table~\ref{tab:multistage_summary} further summarizes the performance of our \multiLFramework~ under human-in-the-loop settings. Both the T1~$\rightarrow$~T2 and L1~$\rightarrow$~L2 configurations correctly handled nearly all incoming samples at the M1 stage (115,789 and 116,669 correctly predicted packets, respectively). Only a few packets required M2-level processing (6 for T1, 4 for L1), and even fewer required human intervention (3 for T1, none for L1). The \textit{human effort} (HE) remained minimal in both cases, just 0.0026\% for the Transformer-based path and 0.0000\% for the LSTM-based path. Due to the higher throughput and lower error rate, we selected the Transformer-based architecture for deployment over the LSTM-based one. The minimal involvement required from human experts proves that our approach is both feasible and effective for real-time scenarios.

\begin{table}[h!]
\centering
\caption{Human Workload Under \multiLFramework~ Using DL-based M1 \& M2}
\resizebox{0.85\columnwidth}{!}{
\begin{threeparttable}
\begin{tabular}{lcc}
\hline
\textbf{Metric} & \textbf{T1 → T2 → Human} & \textbf{L1 → L2 → Human} \\
\hline
\rowcolor{lightblue}
M1 TRP          & 115,795     & 116,673     \\
M1 CPP             & 115,789     & 116,669     \\
\rowcolor{lightblue}
M2 TRP                & 6           & 4           \\
M2 CPP             & 3           & 4           \\
\rowcolor{lightblue}
HIR   & 3           & 0           \\
HE    & 0.0026 \%     & 0.0000 \%        \\
\hline
\end{tabular}
\begin{tablenotes}
\scriptsize
\item
\textbf{TRP}-- Total Received Packets; 
\textbf{CPP}-- Correctly Predicted Packets;
\textbf{HIR}--- Human Intervention Required;
\textbf{HE}--- Human Efforts
\end{tablenotes}
\end{threeparttable}}
\label{tab:multistage_summary}
\end{table}

Further, to demonstrate the significance and sustainability of our proposed approach under the \multiLFramework~framework, we measured the environmental impact and computational cost using the \texttt{CodeCarbon}\footnote{https://codecarbon.io/} library. Table~\ref{tab:environmental_stats} presents the carbon emissions and energy consumption metrics for each deep learning model variant. The results show that the lightweight models (T1 and L1) are notably more energy-efficient and environmentally friendly than their complex counterparts (T2 and L2). For instance, the T1 model consumed only 0.0242 kWh of energy and emitted just 0.0070 kg of CO$_2$, while the L2 model, being the most complex, used 0.1074 kWh and emitted 0.0312 kg of CO$_2$.

\begin{table}[h!]
\centering
\caption{Environmental Impact Statistics of DL-based Models}
\vspace{-2mm}
\resizebox{0.8\columnwidth}{!}{
\begin{threeparttable}
\begin{tabular}{lccccc}
\hline
\textbf{Model} & \textbf{CO$_2$} & \textbf{CPU\_E} & \textbf{RAM\_E} & \textbf{Total\_E} & \textbf{CPU\_P} \\
\hline
\rowcolor{lightblue}
T1 & 0.0070 & 0.0165 & 0.0078 & 0.0242 & 42.5 \\
T2 & 0.0134 & 0.0314 & 0.0148 & 0.0462 & 42.5 \\
\rowcolor{lightblue}
L1  & 0.0158 & 0.0369 & 0.0174 & 0.0543 & 42.5 \\
L2  & 0.0312 & 0.0730 & 0.0343 & 0.1074 & 42.5 \\
\hline
\end{tabular}
\begin{tablenotes}
\scriptsize
\item
\textbf{CO$_2$}-- CO$_2$ Emission (KG); 
\textbf{CPU\_E}-- CPU Energy (kWh);
\textbf{RAM\_E}-- RAM Energy (kWh);
\textbf{Total\_E}--Total Energy (kWh);
\textbf{CPU\_P}-- CPU Power (W)
\end{tablenotes}
\end{threeparttable}}
\label{tab:environmental_stats}
\end{table}

\subsection{Results using Other Benchmark Datasets}
This section presents the performance of the proposed \multiLFramework~framework using two benchmark datasets, M-En Dataset-1 \cite{rustam2024malicious} and M-En Dataset-2 \cite{rustam2024famtds}, which are synthetically created by combining well-known intrusion detection datasets. Dataset-1 is a combination of IoTID-20, UNSW-NB15, and InSDN, while Dataset-2 consists of IoTID-20 and UNSW-NB15. The results show that the M1 correctly classifies the majority of packets, while the M2 effectively handles low-confidence cases passed from M1, as shown in Table \ref{tab:multistage_summary2}. Only a small fraction of packets require human intervention, with human effort reduced to as low as 0.29\% in Dataset-2. These significant results show the effectiveness of the proposed \multiLFramework~in maintaining high accuracy while substantially reducing human workload across different benchmark datasets.

\begin{table}[h!]
\centering
\caption{Proposed \multiLFramework~ Using Benchmark Datasets}
\resizebox{0.65\columnwidth}{!}{
\begin{threeparttable}
\begin{tabular}{lcc}
\hline
\textbf{Metric}~~ & \textbf{Dataset-1 \cite{rustam2024malicious}}~~~~  & \textbf{Dataset-2 \cite{rustam2024famtds}}~~~~~\\
\hline
\rowcolor{lightblue}
M1 TRP          & 20,865     & 22,680     \\
M1 CPP             & 19,428     & 22,481     \\
\rowcolor{lightblue}
M2 TRP                & 1249           & 85           \\
M2 CPP             & 240           & 18           \\
\rowcolor{lightblue}
HIR   & 951           & 67           \\
HE    & 4.5485 \%     &  0.2954\%        \\
\hline
\end{tabular}
\begin{tablenotes}
\scriptsize
\item
\textbf{TRP}-- Total Received Packets; 
\textbf{CPP}-- Correctly Predicted Packets;
\textbf{HIR}--- Human Intervention Required;
\textbf{HE}--- Human Efforts;
\end{tablenotes}
\end{threeparttable}}
\label{tab:multistage_summary2}
\end{table}

\color{black}

\subsection{Comparison With Benchmark Studies}

Many studies validate their malicious traffic detection approaches in offline settings, often achieving significant results. However, these systems frequently experience performance degradation when deployed in real-time environments due to the dynamic nature of network traffic and the emergence of new attack patterns. The datasets used to train these models are typically static, which limits their effectiveness in real-world applications. There are no existing studies that evaluate their work on real-time online M-En traffic. Therefore, for a fair comparison, we implemented existing M-En studies on our newly collected real-time M-En dataset, deploying them according to the methodologies outlined in their respective published work. Since most existing studies conducted only offline testing, we replicated their models in the same offline setting to maintain consistency in evaluation. For instance, Rustam et al.~\cite{rustam2023securing} deployed their approach on an M-En dataset using the S-DATE. While the performance appeared strong after applying the S-DATE data balancing method, the results indicated potential data leakage issues, which could explain the observed high accuracy. Similarly, other studies~\cite{rustam2024malicious,rustam2024famtds} proposed optimization-based approaches to handle diverse traffic patterns in M-En networks, employing PSO and MFO, respectively. These studies introduced PSO-D-SEM and a \emph{Fully Automated Malicious Traffic Detection System} (FAMTDS). Another study~\cite{rustam2024malicious1} utilized a \emph{dual-data trained LightGBM} (DDT-LightGBM) model, achieving significant results in the M-En network within an offline testing framework. Zukaib et al.~\cite{zukaib2024meta} integrated federated learning and meta-learning to propose the meta-fed IDS, which was tested on the M-En dataset. However, their evaluation was limited to offline settings and did not incorporate real-time data collection or adaptation. Furthermore, we deployed modern GNN-based and transformer-based architectures to compare our approach with advanced approaches \cite{10948513,11199889}. In contrast, our study conducted both online and offline testing and was built on a real-time M-En dataset. 

In Table~\ref{tab:MTDS_summary}, we compare our approach, \multiLFramework, to relevant M-En benchmark studies. Most existing approaches~\cite{rustam2023securing,rustam2024malicious,rustam2024famtds,rustam2024malicious1,zukaib2024meta} use static datasets and evaluate performance only in offline scenarios, reporting accuracy values between 0.94 and 0.991. While these systems can be effective in controlled settings, they often lack the mechanisms to continuously learn from new traffic patterns or adapt to real-time fluctuations in M-En networks. In contrast, \multiLFramework~is the only method evaluated in both offline and online modes, achieving near-optimal accuracy scores of 1.00 (offline) and 0.999 (online). These results highlight \multiLFramework's resilience to dynamic traffic behavior. By continuously retraining on fresh data and utilizing multi-level validation checks, \multiLFramework~maintains high accuracy even under unpredictable network conditions, offering a robust solution for real-time malicious traffic detection in M-En environments.

\begin{table}[h!]
    \centering
    \caption{Comparison With Existing Studies}
    \resizebox{0.9\columnwidth}{!}{
        \begin{tabular}{cccccc}
            \hline
            \multirow{2}{*}{ Ref.} &  \multirow{2}{*}{Year} &  \multirow{2}{*}{Approach} & \multicolumn{2}{c}{Testing} & \multirow{2}{*}{Results} \\\cline{4-5}
           &  & & offline & online & \\ \hline
    \rowcolor{lightblue}
             \cite{rustam2023securing} & 2023 & ETC, S-DATE &  $ \checkmark $ & $ \times $ & 0.986\\
            \cite{rustam2024malicious} & 2024 & PSO-D-SEM & $ \checkmark $ & $ \times $  & 0.978 \\
    \rowcolor{lightblue}
            \cite{rustam2024famtds} & 2024 & FAMTDS &  $ \checkmark $ & $ \times $ & 0.991\\
            \cite{rustam2024malicious1} & 2024 & DDT-LightGBM & $ \checkmark $ & $ \times $  & 0.980 \\
    \rowcolor{lightblue}
            \cite{zukaib2024meta} & 2024 & Meta-Fed IDS & $ \checkmark $ & $ \times $ & 0.94 \\
            
            \cite{10948513} & 2025 & GTAE-IDS & $ \checkmark $ & $ \times $  & 0.968 \\
    \rowcolor{lightblue}
            \cite{11199889} & 2025 & GSAGE+RF & $\checkmark $ & $ \times $  & 0.99 \\\hline
            
            Our & 2025 & \multiLFramework~ & $ \checkmark $ & $ \checkmark $ & 1.00, 0.999\\\hline
        \end{tabular}
    }
    \label{tab:MTDS_summary}
\end{table}

\section{Discussion} 

In this study, we collect a dataset in the M-En and propose \multiLFramework~for malicious traffic detection, testing it in both offline and real-time scenarios. We deploy \multiLFramework~to reduce computational costs and enhance accuracy over time. We reduce the computational cost because the initial traffic is evaluated by the lightweight M1 model, which has a faster prediction time due to its lightweight architecture. Most of the traffic is filtered by M1, and only a small portion is forwarded to M2 and, if necessary, to a human expert. Additionally, we improve the approach's performance by incorporating an extra layer of security using M2 and human involvement. 



\begin{figure}[!ht]
    \centering
    \includegraphics[width=\columnwidth]{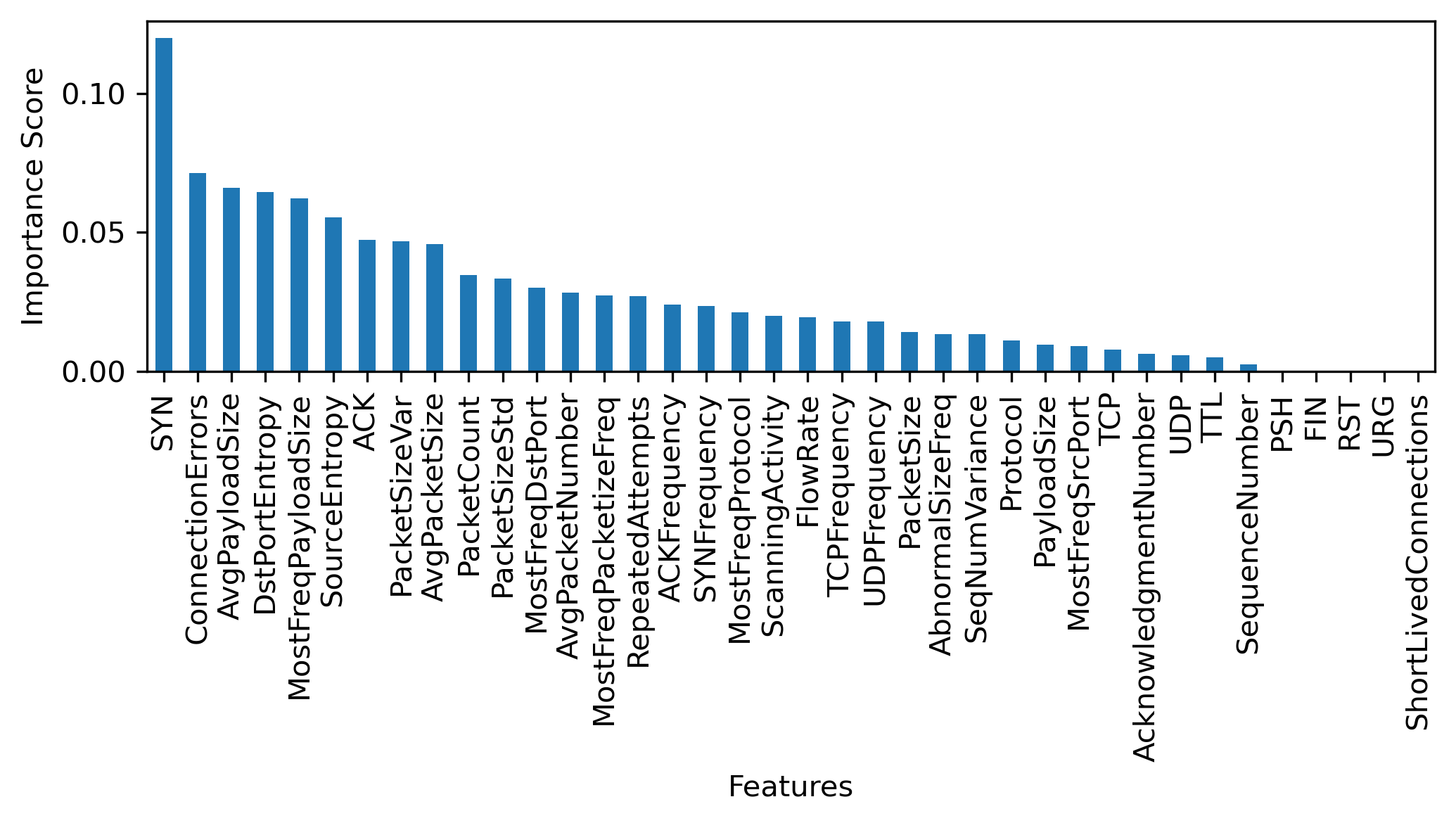}
    \caption{Feature Importance Score}
    \label{fig:FeatImp}
\end{figure}

We achieve continuous improvement in model performance over time through continuous training with new data, enabling the model to efficiently adapt to evolving traffic patterns, as well as through our feature engineering strategy. We utilize general and statistical features, as they both significantly contributed to the model's effectiveness, as illustrated in Figure~\ref{fig:FeatImp}. Statistical features, such as \textit{ConnectionErrors}, \textit{DstPortEntropy}, \textit{MostFreqPayloadSize}, \textit{SourceEntropy}, \textit{MostFreqPacketSizeFreq}, \textit{FlowRate}, \textit{PacketCount}, \textit{PacketSizeVar}, and \textit{AvgPayloadSize}, played a critical role in detecting malicious traffic according to the importance score. For instance, a sudden spike in packet count from a specific IP within a short time window may signal an attack. Additionally, some general features, such as \textit{SYN} and \textit{ACK} flags, are also among the top contributors for malicious traffic detection. These statistical and general features together provided highly correlated input representations, as shown in Figure~\ref{fig:FeatImp}, thereby enhancing the overall performance of the framework.

To justify the need for a robust AI solution for M-En networks and the limitations of existing datasets, we conducted cross-domain experiments. Table~\ref{DDPerf} shows performance variations across scenarios, underscoring the value of a unified M-En-based approach. In our experiments, we first trained the model with IoT data, tested it with traditional (trad.) data, and then trained it on trad. data and tested it with IoT data. The models performed poorly in both cases due to the distinct traffic patterns, demonstrating that existing SOTA datasets are inadequate for creating a unified framework for M-En networks. Subsequently, we conducted experiments using the M-En dataset, training models on it, and testing with both IoT and trad. traffic. The models showed significant improvements in handling both types of traffic, underscoring the effectiveness of our M-En dataset for training models in diverse scenarios.

\begin{table}[h!]
\centering
\caption{Results Using Different Training \& Testing Datasets}\label{DDPerf}
\resizebox{0.9\columnwidth}{!}{
\begin{tabular}{llcccccc}
\hline
Training & Testing & Acc & Class & Pre & Rec & F1  \\ \hline
\rowcolor{lightblue}
IoT & Trad. &
0.006 &
  0 &
  0.01 &
  0.03 &
  0.01 \\
\rowcolor{lightblue}
  & & &
  1 &
  0.00 &
  0.00 &
  0.00  \\

Trad. & IoT &
0.014 &
  0 &
  0.13 &
  0.02 &
  0.03  \\
 &
  & &
  1 &
  0.00 &
  0.00 &
  0.00  \\ 

\rowcolor{lightblue}
M-En & IoT \& Trad. &
0.999 &
  0 &
  1.00 &
  1.00 &
  1.00  \\
\rowcolor{lightblue}
 &
  & &
  1 &
  1.00 &
  1.00 &
  1.00 \\ \hline
\end{tabular}}
\end{table}

\begin{figure}[h!]
\centering
    \begin{subfigure}[b]{\columnwidth}
    \centering
  \includegraphics[width=\columnwidth]{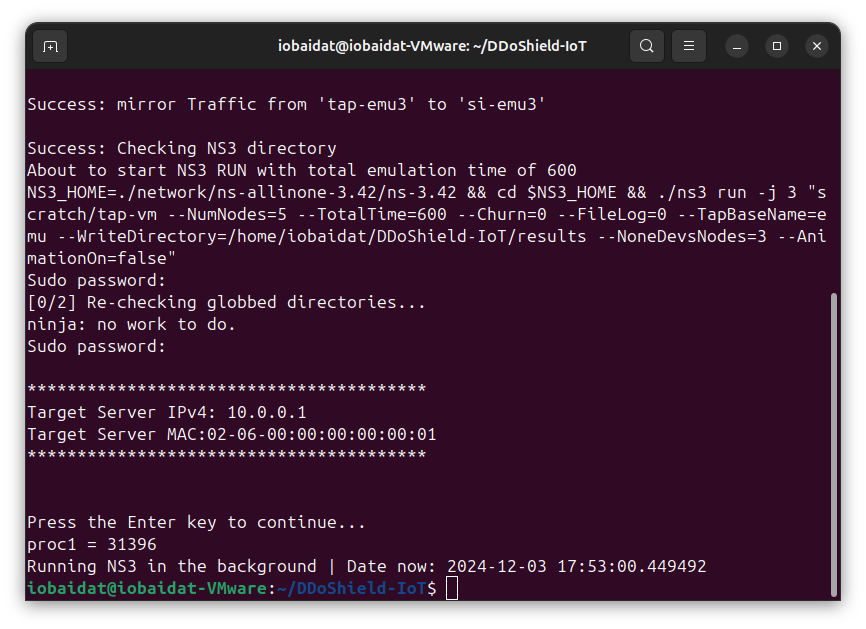}
   \caption{Server Connection Establishment }
    \label{fig1r}
    \end{subfigure}
    \begin{subfigure}[b]{\columnwidth}
    \centering
    \includegraphics[width=\columnwidth]{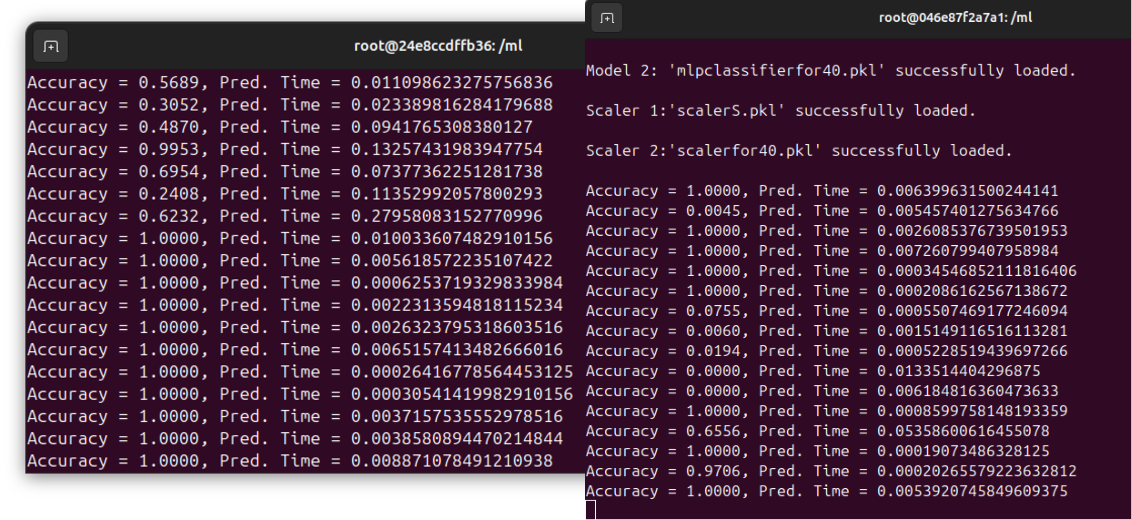}
\caption{Live Performance of Models and Continuous Improvement}    \label{fig2r}
    \end{subfigure}

\caption{Real-time Testing Snapshots }
\label{livereal}
\end{figure}

Furthermore, Figure \ref{livereal} demonstrates the performance of our testing in a real-time scenario. Figure \ref{fig1r} illustrates the connection between the server and the gateway of M-En networks, which facilitates the real-time ingestion of network packets for processing. Once the connection is established, the models begin analyzing the traffic. Initially, the performance is suboptimal, showing accuracy levels around 50-60\%. However, as the system continues to operate, the performance progressively improves due to the continuous learning approach implemented in our framework. The continuous learning mechanism enables the models to adapt to the evolving traffic patterns by retraining on newly collected data in real-time, as shown in Figure \ref{fig2r}. Over time, the accuracy significantly increases, stabilizing at 99-100\%, as seen in the later stages of the testing process. This improvement highlights the robustness and adaptability of our framework, as well as the importance of incorporating real-time data and continual learning for addressing the dynamic nature of M-En network environments. These results further validate the effectiveness of our unified framework and its ability to handle diverse and evolving traffic scenarios in real-time applications.

\textbf{$\Rightarrow$ Significance:} This study has several significances in network security for malicious attack detection and shows a strong contribution in the given domain:
\begin{enumerate*}[label=\textbf{(\textit{\roman*})}]
    \item This study introduces a comprehensive benchmark dataset specifically tailored for M-En networks, providing a foundational resource for researchers to advance security solutions in complex, M-En network settings.
    \item The availability of our research resources in an open repository not only ensures transparency and reproducibility but also encourages further development, validation, and deployment of novel methodologies in the domain of M-En network security.
    \item By incorporating continuous learning and human intervention, the framework demonstrates an ability to adapt dynamically to new and unseen traffic patterns. This adaptability is essential for dealing with zero-day attacks and evolving threats, providing a sustainable approach to long-term network security.
    \item \multiLFramework~ability to leverage a lightweight model (M1) for initial detection significantly reduces computational overhead, which is crucial for real-time applications. This workflow ensures that the framework is not only accurate but also efficient, making it suitable for deployment in resource-constrained environments.
\end{enumerate*}

\textbf{$\Rightarrow$ Limitations:} Despite the significance of this study, there are several limitations to consider:
\begin{enumerate*}[label=\textbf{(\textit{\roman*})}]
    \item The present work focuses on IoT and traditional networks within the M-En framework. This controlled scope ensures methodological clarity and reproducibility. However, the framework is inherently extensible and can seamlessly integrate additional domains, such as SDN and \textit{consumer electronics} (CE) networks, during real-time deployment, enabling broader applicability in larger and more heterogeneous infrastructures. 
    
    \item \multiLFramework incorporates human oversight to validate critical or uncertain detections. While this may introduce latency under high-alert conditions, it enhances interpretability and operational reliability, which are key for security in sensitive environments. Notably, studies and industry reports indicate that a human analyst typically manages 100–200 alerts per day\footnote{https://countershadow.com/blog/ai-vs-cyber-threats-accelerating-response-in-a-24-7-landscape/}, underscoring the value of incorporating guided automation to reduce analyst workload. Future work will focus on integrating active learning and feedback-driven tuning to streamline this process further.
    
    \item The continual learning strategy retrains the model using previously misclassified samples, enhancing adaptability to evolving threats. Although this iterative update process increases computational demand over time, it ensures the model remains robust against concept drift. Future optimization, such as lightweight retraining or selective memory replay, can maintain adaptability while minimizing resource overhead.
\end{enumerate*}

\section{Conclusion \& Future Direction}
\label{sec:coclu_future_work}
In this paper, we first constructed a packet-level M-En dataset (from live \ddosh~traffic mixed with public \pcap~traces) to provide the research community with a realistic corpus and a reproducible pipeline for heterogeneous M-En attack research. Building on this foundation, we presented \multiLFramework, a continuous-learning framework that unifies lightweight and deep architecture models and incorporates selective human feedback to secure M-En networks. 
We deployed \multiLFramework~online in \ddosh~ using a Transformer-based M1 \& M2, maintaining 0.999 accuracy while requiring human intervention to label only 0.0026\% for 115,795 live packets. In contrast, the baseline approach requires significantly more human intervention, 109 expert checks out of 111,644 packets, which represents 0.0976\% of the workload. This online evaluation confirms a high detection rate under real traffic, while the chosen transformer models provide high throughput and consume low power, resulting in 0.007 kg CO$_2$ emission per experiment. 
We concluded that traditional models suffer a drop in accuracy during online testing, whereas our \multiLFramework effectively handles live traffic through layered verification and minimal human intervention, achieving near error-free performance. Additionally, our statistical feature extraction configuration, using a 1-second processing interval, proved effective. Features like \textit{average payload size} and \textit{destination port entropy} significantly improved malicious traffic detection.
Further, our findings show that continuous learning with weight interpolation enables the M1 model to adapt to new traffic and attacks while preserving prior knowledge, minimizing human effort.

In future work, we aim to expand the dataset with encrypted and zero-trust traffic and explore unsupervised drift detection to enable adaptive retraining. We also plan to enrich \ddosh~ by injecting additional binaries and diverse attack types, along with augmenting it using public \pcap~ repositories via our modular pipeline. To evaluate realism, we will compute entropy and protocol-distribution metrics and benchmark them against backbone traces (e.g., MAWI~\cite{cho2000traffic}). Finally, we intend to explore the integration of \textit{Large Language Models} (LLMs) to support or automate human-in-the-loop decisions for improved speed and reliability.

\section*{Data Availability Statement}
The code and datasets used for reproducibility of the results in this study are publicly available on GitHub at \url{https://github.com/furqanrustam/MultiLF}.

\section*{Declaration of competing interest}
The authors declare that they have no known competing financial interests or personal relationships that could have appeared to influence the work reported in this paper.

\section*{CRediT authorship contribution statement}
\textbf{Furqan Rustam:} Conceptualization, Data curation, Formal analysis, Validation, Investigation, Methodology, Software, Validation, Visualization, Writing – original draft.  
\textbf{Islam Obaidat:} Data curation, Formal analysis, Investigation, Validation,  Resources, Writing – review \& editing.  
\textbf{Anca Delia Jurcut:} Formal analysis, Investigation, Project administration, Resources, Supervision, Writing – review \& editing.

\section*{Acknowledgment}
This work is funded by the School of Computer Science, CHIST-ERA ERA-NET - SPiDDS Topic, and the Irish Research Council (IRC).


\bibliography{02-references}
\bibliographystyle{unsrt}

\newpage
\appendix
\onecolumn\section{Statistical Feature Definitions and Calculation Details}
\label{app1}

\begin{tcolorbox}[colback=white, colframe=black!60, title=Statistical Feature, fonttitle=\bfseries,fontupper=\ttfamily\scriptsize ]
\begin{enumerate}
    \item \textbf{\textit{Packet Count}}: The total number of packets observed in each time window, represented as \( \text{PacketCount}_{t} = \sum_{i=1}^{n} \text{Packet}_{i} \).
    \item \textbf{\textit{Destination Port Entropy}}: Measures the entropy of destination ports to detect scanning activities, defined as \( \text{Entropy} = -\sum_{i=1}^{n} p_{i} \log(p_{i}) \).
    \item \textbf{\textit{Most Frequent Source Port}}: Identifies the most common source port in a window, \( \text{SourcePort}_{\max} \).
    \item \textbf{\textit{Most Frequent Destination Port}}: Identifies the most common destination port in a window, \( \text{DestinationPort}_{\max} \).
    \item \textbf{\textit{Short-lived Connections}}: Counts the number of short-lived connections, \( \text{ShortLivedConnections} = \sum_{i=1}^{n} \delta_{i} \), where \( \delta_{i} \) indicates a short-lived connection.
    \item \textbf{\textit{Repeated Connection Attempts}}: Measures repeated connection attempts, \( \text{RepeatedAttempts} = \sum_{i=1}^{n} \alpha_{i} \), where \( \alpha_{i} \) indicates a repeated attempt.
    \item \textbf{\textit{Network Scanning Activity}}: Counts instances of SYN flags without ACK flags, \( \text{Scanning} = \sum_{i=1}^{n} (\text{SYN} - \text{ACK}) \).
    \item \textbf{\textit{Flow Rate}}: Calculated as packets per second, \( \text{FlowRate} = \frac{\text{TotalPackets}}{\text{TimeInterval}} \).
    \item \textbf{\textit{Source Entropy}}: Entropy of source addresses, \( \text{SourceEntropy} = -\sum_{i=1}^{n} q_{i} \log(q_{i}) \), where \( q_{i} \) is the probability distribution of source addresses.
    \item \textbf{\textit{Connection Errors (RST flag)}}: Counts instances of RST flags, \( \text{RSTCount} = \sum_{i=1}^{n} \text{RST}_{i} \).
    \item \textbf{\textit{Most Frequent Packet Size Frequency}}: Identifies the most common packet size, \( \text{PacketSize}_{\max} \).
    \item \textbf{\textit{Abnormal Size Frequency}}: Counts packets exceeding a size threshold, \( \text{AbnormalSizeFrequency} = \sum_{i=1}^{n} \text{Packet}_{i} \text{ if size}_{i} > \text{Threshold} \).
    \item \textbf{\textit{Sequence Number Variance}}: Variance in sequence numbers, \( \text{Var}( \text{SequenceNumber}) \).
    \item \textbf{\textit{Average Packet Number}}: Average packets per interval, \( \text{AvgPackets} = \frac{\text{TotalPackets}}{\text{TimeIntervals}} \).
    \item \textbf{\textit{SYN Frequency}}: Frequency of SYN flags, \( \text{SYNFrequency} = \frac{\text{TotalSYN}}{\text{TimeInterval}} \).
    \item \textbf{\textit{ACK Frequency}}: Frequency of ACK flags, \( \text{ACKFrequency} = \frac{\text{TotalACK}}{\text{TimeInterval}} \).
    \item \textbf{\textit{TCP Frequency}}: Proportion of TCP packets, \( \text{TCPFrequency} = \frac{\text{TotalTCP}}{\text{TotalPackets}} \).
    \item \textbf{\textit{UDP Frequency}}: Proportion of UDP packets, \( \text{UDPFrequency} = \frac{\text{TotalUDP}}{\text{TotalPackets}} \).
    \item \textbf{M\textit{ost Frequent Protocol}}: Most used protocol, \( \text{Protocol}_{\max} \).
    \item \textbf{\textit{Packet Size Variability}}: Variance in packet sizes, \( \text{Var}( \text{PacketSize}) \).
    \item \textbf{\textit{Most Frequent Payload Size}}: Most common payload size, \( \text{PayloadSize}_{\max} \).
    \item \textbf{\textit{Average Payload Size}}: Mean payload size, \( \text{AvgPayloadSize} = \frac{\text{TotalPayload}}{\text{TotalPackets}} \).
    \item \textbf{\textit{Packet Size Standard Deviation}}: Standard deviation of packet sizes, \( \text{StdDev}( \text{PacketSize}) \).
    \item \textbf{\textit{Average Packet Size}}: Mean size of packets within a window, \( \text{AvgPacketSize} = \frac{\text{TotalPacketSize}}{\text{TotalPackets}} \).
\end{enumerate}
\end{tcolorbox}


\end{document}